\documentclass[journal,twoside,web]{ieeecolor}
\usepackage{tmi}
\usepackage{cite}
\usepackage{multirow}
\usepackage{diagbox}
\usepackage{slashbox}
\usepackage{amsmath,amssymb,amsfonts}
\usepackage{graphicx}
\usepackage{textcomp}
\usepackage{amsmath}
\usepackage{amssymb}
\usepackage{algorithm}
\usepackage{algcompatible}
\usepackage{varwidth}
\usepackage{subcaption}
\algnewcommand\INPUT{\item[\textbf{Input:}]}
\algnewcommand\OUTPUT{\item[\textbf{Output:}]}

\def\YZ#1{{\color{black} {\bf}{{#1}}{\bf{}}}}

\def\BibTeX{{\rm B\kern-.05em{\sc i\kern-.025em b}\kern-.08em
    T\kern-.1667em\lower.7ex\hbox{E}\kern-.125emX}}
\begin{document}
\title{Early Melanoma Diagnosis with Sequential Dermoscopic Images}
\author{Zhen Yu, Jennifer Nguyen, Toan D Nguyen, John Kelly, Catriona Mclean, Paul Bonnington, Lei Zhang, Victoria Mar, and Zongyuan Ge, \IEEEmembership{Member, IEEE}
\thanks{The authors acknowledge support from NHMRC Centre of Research Excellence in Melanoma (1135285), National Natural Science Foundation of China (81950410639), Outstanding Young Scholars Support Program (3111500001), Xi’an Jiaotong University Basic Research and Profession Grant (xtr022019003, xzy032020032), Epidemiology modeling and risk assessment (20200344) and Xi’an Jiaotong University Young Scholar Support Grant (YX6J004).}
\thanks{Z. Yu, and L. Zhang are with the Central Clinical School, Faculty of Medicine, Nursing and Health Sciences, Monash University, Clayton, VIC 3800, Australia (E-mail: zhen.yu@monash.edu, lei.zhang1@monash.edu).}
\thanks{L. Zhang is also with the China-Australia Joint Research Center for Infectious Diseases, School of Public Health, Xi’an Jiaotong University Health Science Center, Xi’an, Shaanxi, 710061, China;
Artificial Intelligence and Modelling in Epidemiology Program, Melbourne Sexual Health Centre, Alfred Health, Melbourne, Australia; and 
Department of Epidemiology and Biostatistics, College of Public Health, Zhengzhou University, Zhengzhou 450001, Henan, China.}
\thanks{J. Nguyen, J. Kelly, C. Mclean and V. Mar are with the Victorian Melanoma Service, Alfred Health, Melbourne, VIC 3004, Australia (Email: je.nguyen@alfred.org.au, victoria.mar@monash.edu)}
\thanks{Toan D Nguyen, P. Bonnington and Zongyuan. Ge are with the eResearch Centre, Monash University, Clayton, VIC 3800, Australia (e-mail: Paul.Bonnington@monash.edu).}
\thanks{Zongyuan. Ge is also with the Monash Medical AI and Monash Airdoc Research Centre in Monash University, Clayton, VIC 3800, Australia (e-mail: zongyuan.ge@monash.edu).}}
\maketitle

\begin{abstract}
Dermatologists often diagnose or rule out early melanoma by evaluating the follow-up dermoscopic images of skin lesions. However, existing algorithms for early melanoma diagnosis are developed using single time-point images of lesions. Ignoring the temporal, morphological changes of lesions can lead to misdiagnosis in borderline cases. In this study, we propose a framework for automated early melanoma diagnosis using sequential dermoscopic images. To this end, we construct our method in three steps. First, we align sequential dermoscopic images of skin lesions using estimated Euclidean transformations, extract the lesion growth region by computing image differences among the consecutive images, and then propose a spatio-temporal network to capture the dermoscopic changes from aligned lesion images and the corresponding difference images. Finally, we develop an early diagnosis module to compute probability scores of malignancy for lesion images over time. We collected 179 serial dermoscopic imaging data from 122 patients to verify our method. Extensive experiments show that the proposed model outperforms other commonly used sequence models. We also compared the diagnostic results of our model with those of seven experienced dermatologists and five registrars. Our model achieved higher diagnostic accuracy than clinicians (63.69\% vs. 54.33\%, respectively) and provided an earlier diagnosis of melanoma (60.7\% vs. 32.7\% of melanoma correctly diagnosed on the first follow-up images). These results demonstrate that our model can be used to identify melanocytic lesions that are at high-risk of malignant transformation earlier in the disease process and thereby redefine what is possible in the early detection of melanoma.
\end{abstract}

\begin{IEEEkeywords}
Lesion alignment, sequential dermoscopic images, spatio-temporal feature learning, early melanoma diagnosis
\end{IEEEkeywords}

\section{Introduction}
\begin{figure}[h]
\centering
\includegraphics[width=0.47\textwidth]{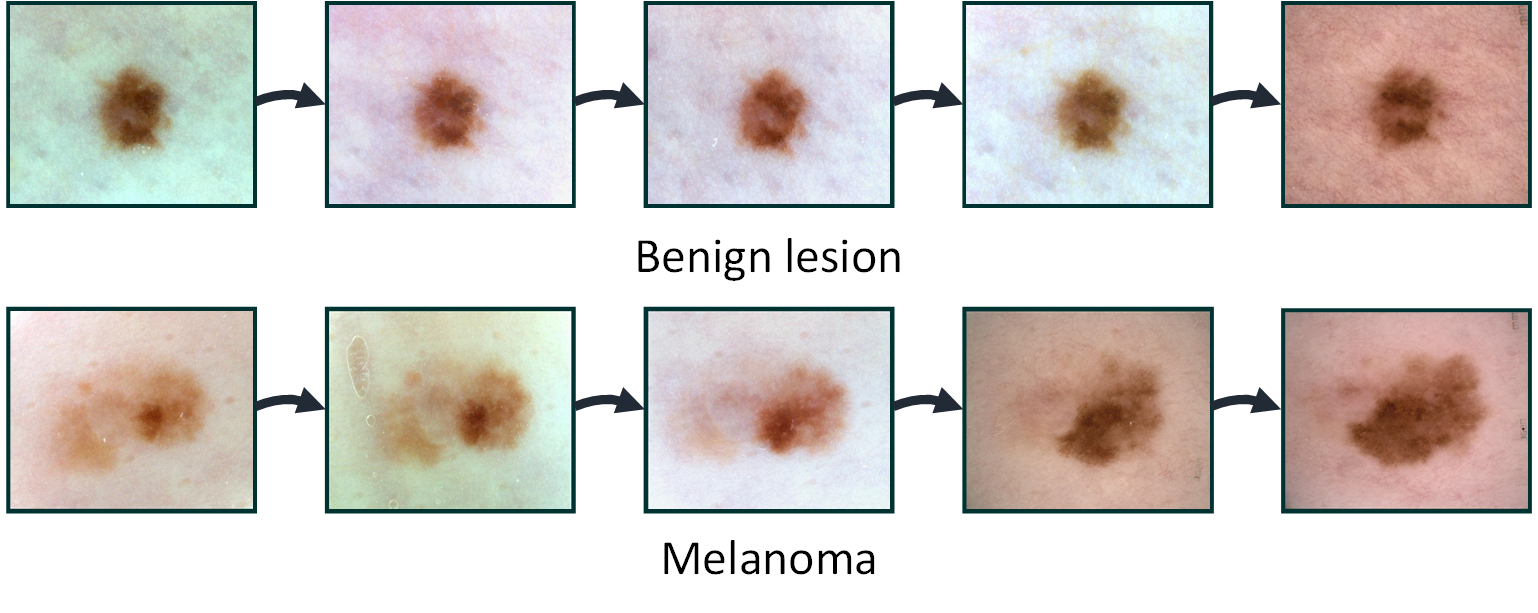}
\caption{Lesions de facto are progressively evolving. The benign lesion remains fairly stable in terms of colour and shape, whereas the malignant melanoma exhibits substantial focal enlargement.}
\label{fig: evo}
\end{figure}

\IEEEPARstart{E}{arly} diagnosis of malignant melanoma is crucial, as patients can be cured of the melanoma by surgically excising the primary tumour during early, non-invasive stages. For decades, visual dermoscopic examinations have been widely adopted for recognizing melanoma. Existing criteria, such as the `7-point checklist', enable accurate identification of melanoma with distinct dermoscopic features \cite{rigel2010evolution,mackie1989malignant,abbasi2004early,schadendorf2015melanoma}. However, melanoma at an early stage may be subtle and lacks the dermoscopic criteria for malignancy (e.g., asymmetric shape, irregular pigment network, or diverse pigmentation), making early diagnosis challenging \cite{salerni2012benefits, malvehy2002follow}. In \cite{rosendahl2012impact}, the number of lesions biopsied to find a single melanoma (“number needed to treat”, or NNT) varied from 8.9 14.6. Thus, in practice, there is a delicate balance between the failure to recognise melanoma and the clinical over-diagnosis of benign melanocytic nevi. Therefore, dermoscopic monitoring has been proposed to monitor inconspicuous lesions, and lesion evolution is an additional criterion used to improve the diagnostic accuracy of borderline lesions \cite{abbasi2004early, rigel2010evolution}. The rationale behind this is that benign melanocytic naevi will remain fairly stable over time, whereas melanoma may develop from a pre-existing benign naevus, with changes diagnostic for melanoma difficult to appreciate at a single time point \YZ{(see Fig. \ref{fig: evo})}. Studies \cite{rajgopal2017dangers, pampena2018nevus, haenssle2016association} have shown that approximately 30\%$\sim$50\% of melanomas arise from pre-existing benign lesions. Accumulating evidence suggests that evaluation of lesion changes with sequential dermoscopic images will substantially enhance the ability of clinicians to recognize melanomas at earlier stages \cite{kittler2006identification, abbasi2004early, rigel2010evolution, moloney2014detection}. Nevertheless, visually differentiating early melanoma from benign lesions remains a challenge because of the subjectivity of human cognitive functioning and variation in the experience of clinicians.  

Artificial intelligence (AI) algorithms have recently demonstrated remarkable performance in dermatology. Deep learning-based techniques are the most promising, and have achieved performance at least equivalent to that of experienced clinicians in image-based diagnosis under experimental conditions \YZ{\cite{yu2016automated, esteva2017dermatologist, brinker2019deep, ge2017skin, moloney2014detection, haenssle2020man}}. Esteva et al. \cite{esteva2017dermatologist} achieved dermatologist-level diagnostic accuracy of melanoma classification by training a deep convolutional neural network (CNN) with more than 120,000 single time-point clinical images. Brinker et al. \cite{brinker2019deep} validated a CNN model for recognizing malignant melanoma on $\sim$12,000 publicly available dermoscopic images, and their model outperformed 136 of 157 dermatologists. Most existing deep learning models output probability scores of lesions to be diagnosed as melanoma, with little information on how the diagnosis was reached. There is concern that this `black box' effect could lead dermatologists astray. As a result, researchers have also explored the use of CNN models to detect dermoscopic features \cite{li2018evidence, kawahara2018fully} and imitate the dermoscopic criteria for diagnosis of melanoma to provide more explainable diagnostic results \cite{gonzalez2018dermaknet, kawahara2018seven}. Although these studies show great potential to improve melanoma diagnosis, their algorithms all use single time-point images. The static nature of lesion presentation can be problematic in early melanoma recognition, for algorithms and clinicians alike. In the context of incipient melanoma, achieving early recognition requires consideration of subtle lesion changes over time. Hence, it is essential to model lesion evolution with sequential dermoscopic images to improve diagnostic algorithms and tools for the surveillance of high-risk individuals.

Several studies have developed algorithms for sequential dermoscopic image analysis \cite{huang2007new, anagnostopoulos2013image, maglogiannis2003automated, li2016skin}, yet their main focuses were on skin lesion image registration or lesion tracking, without assessing lesion evolution for early melanoma diagnosis. Navarro et al. \cite{navarro2018accurate} directly computed the pixel value difference between registered skin lesion image pairs and measured the evolution of the lesion’s size. Although they considered changes in lesion diameter, they did not evaluate the evolution of dermoscopic features for subsequent melanoma diagnosis. Moreover, the dataset used in their study was small (10 image pairs across months rather than years). Very recently, Zhang et al. \cite{boyanzhang} proposed a Siamese neural network to detect short-term lesion changes from dermoscopic image pairs by simply giving predictions of `changed' or `unchanged', however the results of the lesion changes were not further assessed for diagnosis. 

In this study, we propose to model lesion evolution with sequential dermoscopic images for early melanoma diagnosis. Our goal is to incorporate the temporal dynamics of lesion changes as an additional clue and thereby improve diagnostic accuracy of melanoma recognition at an early stage. To this end, we formulate our framework in three steps: 1) skin lesion image alignment, 2) spatio-temporal feature learning, and 3) classification for early diagnosis. We first align lesion images at different time points into the same coordinates, and extract lesion modification regions by computing pixel-level differences between consecutive images. Then, we adopt a two-stream network to learn spatio-temporal features from the aligned dermoscopic images, as well as from difference images to capture subtle lesion changes. Finally, we train the classifier on the aggregated spatio-temporal features and output predictions for each lesion at individual time points to achieve diagnosis earlier in the image sequence. \YZ{Both our problem setting on early melanoma diagnosis using serial data and the proposed framework to register lesion and incorporate the relevant spatio-temporal information are new in the community.} 

The main contributions of this study are summarized as follows:
\begin{itemize}
  \item We develop an image-based AI for early melanoma recognition with clues from lesion evolution, instead of relying solely on the static presentation of lesions. To the best of our knowledge, this is the first study to model lesion modification using serial dermoscopic images for early melanoma diagnosis.
   
  \item We collect a dataset consisting of histologically confirmed serial images of 179 individual skin lesions to evaluate the effectiveness of the proposed approach. Experimental results demonstrate the benefit of algorithm development using serial images compared with that of using static images and demonstrate the superiority of our method over other sequential learning models.
  
  \item We invite 12 dermatologists and dermatology registrars to assess our serial skin lesion image dataset, and further compare their performance with the diagnostic results from our model. The in-depth analysis of the results provides some experimental evidence that the proposed model may achieve earlier and more accurate diagnoses than clinicians.
\end{itemize}

\section{Related works}
\subsection{Skin Lesion Image Alignment}
Image alignment, also known as image registration, is the process of finding the correspondence between two images and transforming the two images into the same coordinate system with matched contents \cite{zitova2003image}. Aligning skin lesion images enables comparison of lesions at different times to evaluate lesion changes. Although image registration has been extensively explored in various medical image analysis tasks \cite{haskins2020deep, hill2001medical}, only a few researchers attempted to align images of skin lesions.

Ilias \cite{maglogiannis2003automated} proposed a hybrid algorithm that aligns dermatological images by separately searching four parameters of geometric transformation with log-polar transformation and a predefined similarity criterion. Anagnostopoulos et al. \cite{anagnostopoulos2013image} utilized a modified scale invariant feature transform (SIFT) with random sample consensus (RANSAC) to estimate affine transformations for the registration of dermoscopic images. Huang et al. \cite{huang2007new} treated melanoma registration as a bipartite graph matching problem and computed a bipartite graph from segmented lesion images. Recently, Li et al. \cite{li2016skin} explored the detection and tracking of lesions from total body images using a deep neural network. Fulgencio et al. \cite{navarro2018accurate} combined superpixel and SIFT descriptors to detect and describe local points in dermoscopic image pairs, and then aligned images using the estimated geometric transformation from matched features. In these studies, however, researchers either stopped at the stage of skin lesion registration without further exploring melanoma diagnosis with the aligned lesion images or simply performed registration on synthetic lesion images. In contrast, in the present study, we align follow-up images of skins and further study the modelling of lesion evolution with aligned serial lesion images for early melanoma diagnosis. 

\subsection{Sequential Images Modelling}
The core purpose of modelling sequential images is to effectively learn discriminative spatio-temporal features. Existing methods for modelling sequential images can be largely grouped into three categories. Methods in the first category usually utilize convolutional neural networks (CNNs) to extract high-level abstract representations from each input image, and then perform temporal aggregation via general pooling or recurrent neural networks (RNNs)~\cite{yue2015beyond,wang2016temporal}. Although this type of method forms a popular baseline in modelling temporal relations from serial data, high-level CNN features lack detailed spatial information and are therefore not suitable for capturing subtle dermoscopic changes. The second category constitutes approaches that directly learn spatio-temporal features using 3D networks or pseudo 3D networks \cite{tran2018closer,xie2018rethinking,sun2015human}. By stacking multiple images as inputs, these models can effectively extract the discriminative spatio-temporal features. State-of-the-art results were achieved in a range of sequential image learning tasks, especially in the video analysis domain \cite{tran2018closer,xie2018rethinking}. Nevertheless, in the medical domain, serial dermoscopic imaging samples from each patient vary in length and often have much fewer images when compared to the available frames from video sequences (e.g., 3$\sim$5/patient vs. more than 30/video clip). \YZ{Hence, it is impractical to directly design a 3D network or use a pre-trained 3D network on our sequential skin lesion imaging data \cite{varol2017long,tran2018closer}}. The third category of approaches \cite{simonyan2014two,feichtenhofer2016convolutional,ng2018temporal} decompose spatio-temporal feature learning tasks by explicitly learning spatial characterizations and temporal evolutions with two-stream network architectures. Generally, the spatial stream accepts RGB images as input, whereas the temporal stream accepts optical flow or RGB difference images as input. Such two-stream networks have been demonstrated to be very effective in learning temporal relations from sequential images, and achieved competitive results compared with that of a heavy 3D CNN. In our study, we designed a two-stream network for modelling lesion changes, but we further connected the two sub-networks with multiple feature difference extraction modules. Accordingly, our model is capable of learning spatio-temporal dermoscopic features from both pixel-level differences and differential CNN feature maps.

\subsection{Computer-aided Early Melanoma Diagnosis}
Over the past several years, a large number of algorithms have been proposed for automated melanoma diagnosis \cite{barata2018survey, pacheco2019recent, pathan2018techniques}. An overwhelming majority of them use deep CNNs as their backbones due to recent advancements in deep learning techniques \cite{lecun2015deep,sun2019optimization, minar2018recent} and the release of publicly available skin lesion datasets \cite{gutman2016skin, tschandl2018ham10000}. Yu et al. \cite{yu2016automated} presented a very deep CNN and a set of schemes to classify melanomas using limited training data. Mobiny et al. \cite{mobiny2019risk} proposed a Bayesian network for recognizing skin lesion cancer. Esteva et al. \cite{esteva2017dermatologist} fine-tuned a CNN model with more than 120,000 images and achieved dermatologist-level diagnostic performance. Studies \cite{haenssle2016association,brinker2019deep,tschandl2019expert} presented CNN models that either outperform or on par with dermatologists. Other efforts have been made to identify skin cancer using algorithms such as ensembles of different models \cite{codella2017deep,gessert2020skin}, feature aggregation \cite{yu2018melanoma,yu2020convolutional}, multi-stage CNN models \cite{xie2020mutual,nida2019melanoma}, and a combination of multimodal data \cite{liu2020deep,gessert2020skin}. \YZ{In addition, a new deep CNN that combined dermoscopic data with clinical data (e.g., age, sex, diameter, and body location of lesion) was developed for the subtle differential diagnosis of early melanomas from their simulator’s dysplastic nevi \cite{tognetti2021new}}. Because deep learning models are usually considered to be uninterpretable and dermatologists are concerned with how CNN models provide predictions \cite{zakhem2018should}, several studies have explored constructing models in a more intuitive manner. Kawahara et al.\cite{kawahara2018fully} proposed a model for detecting dermoscopic features for melanoma recognition. In \cite{kawahara2018seven}, the author developed an algorithm directly modelling the dermoscopic criteria of the 7-point checklist and providing prediction for each criterion respectively. However, all existing studies on computer-aided melanoma diagnosis were designed for identifying cancerous melanoma from other types of lesions using single time-point images. In these studies, the researchers ignored the diagnostic performance of their models on incipient melanoma or featureless lesions. In contrast, in this study, we incorporate clues of lesion changes with sequential dermoscopic images to recognize malignant melanomas early in their evolution. To the best of our knowledge, this is the first study on an algorithm for early melanoma diagnosis using serial dermoscopic imaging data. \YZ{Additionally, extensive experiments verified that the proposed model is capable of outperforming other commonly used sequence models and achieving earlier and more accurate performance than clinicians.}

\section{Method}
\begin{figure}[ht]
\centering
\includegraphics[width=0.48\textwidth]{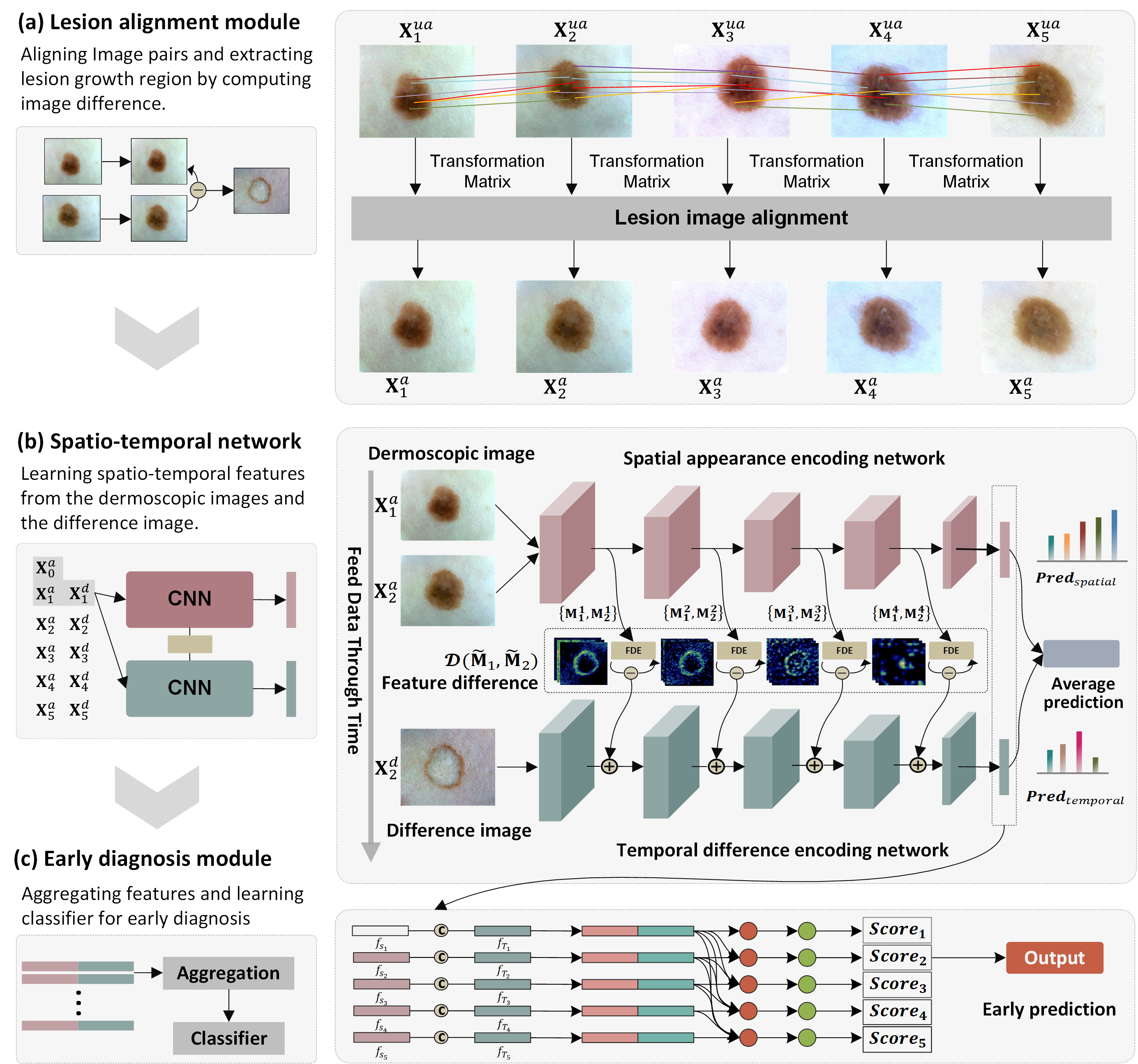}
\caption{ Overview of the proposed method for early melanoma diagnosis. (a) shows the lesion image alignment module. (b) shows the architecture of the spatio-temporal network (STN). (c) is the early diagnosis module.}
\label{fig: framework}
\end{figure}

An overview of the proposed method is presented in Fig. \ref{fig: framework}. The proposed method includes three key components. The \textbf{lesion alignment module}, which aligns lesion images at different time points into the same coordinate system to extract the lesion growth region; the \YZ{\textbf{spatio-temporal network}}, which learns spatio-temporal features from aligned sequential images using an interconnected two-stream network; and the \textbf{early diagnosis module}, which achieves early melanoma diagnosis with the learned spatio-temporal feature using a sequential-based contextual aggregation module and a knowledge distillation training strategy.

\subsection{Skin Lesion Alignment Module}
Because a lesion may vary in its viewpoint and location when images are captured at different time points, we first apply an alignment module to offset these variations. The alignment also assists in tracking changes in dermoscopic features over time. The alignment module consists of three stages: local feature detection, feature matching, and image transformation. In contrast to existing studies which prefer aligning lesion images using a similarity transformation \cite{anagnostopoulos2013image, navarro2018accurate}, we instead use a rigid transformation (also known as Euclidean transformation), because scaling a lesion image will distort the measurement of actual lesion enlargement statistics. 

We define the unaligned image sequence of the $\textit{i}$-th lesion with $\small{}\textbf{N}$ screenings as $\small{\mbox{\textbf{X}}_{i}^{ua} = \left \{ \textbf{X}_{i1}^{ua}, \textbf{X}_{i2}^{ua}, ...,
\textbf{X}_{iN}^{ua}\right\}}$\footnote{$\textit{i}$ will be omitted in the following sections for simplicity.}. For each dermoscopic image sequence $\small{\mbox{\textbf{X}}}^{ua}$, we use the first image as the reference image and perform image alignment sequentially from the second image until the last image. We resize all images to a fixed size of 400$\times$320\footnote{To avoid distorting the shape of lesions, we resize all images with a short side length of 320 while maintaining the aspect ratio. For subsequent feature learning, we crop aligned images with a size of 320$\times$320 as input.} and then detect local key points from each image pair using the accelerated KAZE algorithm (AKAZE)\cite{alcantarilla2011fast}. Subsequently, we perform feature matching among the key points of the image pairs using the Hamming distance (HD) and calculate a $3\times3$ transformation matrix for alignment using random sample consensus (RANSAC). Finally, we denote the aligned image sequence as $\small{\mbox{\textbf{X}}^{a} = \left \{ \textbf{X}_{1}^{a}, \textbf{X}_{2}^{a}, ..., \textbf{X}_{N}^{a}\right\}}$. We summarize the detailed implementation of the lesion alignment module in Algorithm \ref{ag: align}.

\begin{algorithm}[ht]
\scriptsize
\caption{Skin lesion image alignment with rigid transformation.}
\begin{algorithmic}[1]
\INPUT Skin lesion image sequence $\scriptsize{\mbox{\textbf{X}}^{ua} = \left \{ \textbf{X}_{1}^{ua},
\textbf{X}_{2}^{ua}, ...,
\textbf{X}_{N}^{ua}\right\}}$, threshold of the key point detector $\boldsymbol{\theta}_{d}$

\OUTPUT Aligned image sequence $\scriptsize{\mbox{\textbf{X}}^{a} = \left \{ \textbf{X}_{1}^{a},
\textbf{X}_{2}^{a}, ...,
\textbf{X}_{N}^{a}\right\}}$

\STATE \textbf{Initialization:} Set reference image $\textbf{X}_{ref} = \textbf{X}_{1}^{a} = Image\_resize \left (\textbf{X}_{1}^{ua}\right)$, and note the image to be aligned as $\textbf{X}_{dst}$

\STATE \textbf{for} \textit{i} in $\left \{ 2, ..., N \right \}$ \textbf{do:}
\STATE \ \ $\textbf{X}_{dst} = Image\_resize \left (\textbf{X}_{1}^{ua} \right)$
\STATE \ \ $\boldsymbol{kp}_{ref}, \ \boldsymbol{des}_{ref} = AKAZE \left (\textbf{X}_{ref}, \boldsymbol{\theta}_{d} \right)$

\STATE \ \ $\boldsymbol{kp}_{moving}, \ \boldsymbol{des}_{dst} = AKAZE \left (\textbf{X}_{dst},\ \boldsymbol{\theta}_{d} \right)$

\STATE \ \ $\boldsymbol{matches} = HD\_Matching\left (\boldsymbol{des}_{ref}, \  \boldsymbol{des}_{dst} \right)$

\STATE \ \ $\boldsymbol{kp}'_{ref},\ \boldsymbol{kp}'_{dst} = \small{Convert} \left( \boldsymbol{matches},\ \boldsymbol{kp}_{ref},\  \boldsymbol{kp}_{dst} \right)$

\STATE \ \ Compute transformation matrix: $\boldsymbol{T} = RANSAC \left( \boldsymbol{kp}'_{ref}, \ \boldsymbol{kp}'_{dst} \right)$

\STATE \ \ $\textbf{X}'_{dst} = Image\_warping \left( \textbf{X}_{dst}, \boldsymbol{T} \right)$

\STATE \ \ $\textbf{X}_{ref} = \textbf{X}_{i}^{a} = \textbf{X}'_{dst}$

\STATE \textbf{return} $\left \{ \textbf{X}_{1}^{a},
\textbf{X}_{2}^{a}, ...,
\textbf{X}_{N}^{a}\right \}$
\end{algorithmic}
\label{ag: align}
\end{algorithm}

\subsection{Spatio-Temporal Feature Learning Module}
The spatio-temporal feature learning module consists of two sub-networks: a spatial appearance encoding network and a temporal difference encoding network. The \YZ{spatio-temporal network} aims to simultaneously learn abstract appearance representations from individual lesions while also capturing the temporal relations between consecutive images from both raw image pixel differences and multi-level CNN feature differences.
\subsubsection{Spatial Appearance Encoding Network} The spatial network is utilized to encode dermoscopic images into different levels of appearance abstraction which will be incorporated into the temporal network. We employ an off-the-shelf ImageNet pre-trained ResNet-34~\cite{he2016deep} as the backbone. The output of the spatial network is obtained by averaging the prediction scores of individual lesions from the input sequence: 
\begin{equation}
\footnotesize
{\textbf{\textit{Pred}}_{spatial} =  \frac{1}{N}\sum_{t=1}^{N}\mathcal{P_{S}}\left(\mbox{\textbf{X}}_{t}^{a}, \mbox{\textbf{W}}_{S}\right)}
\end{equation}
where $\scriptsize{\mathcal{P_{S}}\left({\mbox{\textbf{X}}}_{t}^{a}, \mbox{\textbf{W}}\right)}$ denotes the mathematical expression of the spatial network with parameters $\small{\mbox{\textbf{W}}_{S}}$ that operates on one dermoscopic image of $\small{\mbox{\textbf{X}}}_{t}^{a}$.

\subsubsection{Temporal Difference Encoding Network} Similar to the spatial network, we use the ResNet-34 as the backbone of the temporal network. Instead of providing the network with static inputs, we provide the difference in pixel intensities between two consecutive images into the temporal network. For each image sequence $\small{}\textbf{X}^{a}$, the image differential map $\small{}{\mbox{\textbf{X}}}_{t}^{d}$ at time $\textit{t}$ is defined as the pixel-wise value subtraction between consecutive dermoscopic images of $\small{}\textbf{X}_{t-1}^{a} $and $\small{}\textbf{X}_{t}^{a}$. To suppress noise from irrelevant contexts, we implement the colour constancy algorithm $\small{\mathcal{C} \left( \cdot \right)}$ based on the general Gray World \cite{van2007edge} and a hair removal function  $\small{\mathcal{H} \left( \cdot \right)}$ \YZ{which was realized by contour detection and morphological filtering}: 
\begin{equation}
\footnotesize
{\textbf{\mbox{X}}}_{t}^{d} =  \mathcal{C} \left ( \mathcal{{H}} \left( \mbox{\textbf{X}}_{t}^{a} \right ) \right),
- \mathcal{C} \left ( \mathcal{{H}} \left( \mbox{\textbf{X}}_{t-1}^{a} \right)
 \right)
\end{equation}
Our motivation is that subtle dermoscopic changes can be directly reflected by pixel distinctions after image alignment. As shown in Fig. 2 (b), the differential image clearly exhibits enlargement of the lesion, which is one of the key malignant features for melanoma diagnosis. Thus, we can explicitly learn the temporal evolution of lesions from pixel-level modifications at this branch.

\subsubsection{Feature Difference Extraction} In contrast to computing skin lesion difference in the raw pixel space, the abstract appearances captured by the CNN are more robust to translation and condition changes \cite{feichtenhofer2016convolutional, ng2018temporal}. Hence, we further incorporate spatial feature differential information from consecutive images into the temporal encoding network by adding the feature map element-wise to the corresponding layers. 

Specifically, during the forward passing of a dermoscopic image sequence, we insert the \YZ{feature difference extraction block (FDE)} at each stage of the spatial encoding network to extract multiple levels of spatial differential features between consecutive images. At time $\textit{t}$ for each image sequence, we have:
\begin{equation}
\footnotesize
\mathcal{D}\left(\widetilde{\mbox{\textbf{M}}}_{t-1}, \widetilde{{\mbox{\textbf{M}}}}_{t}\right) 
= \left \{  
\left( \mbox{\textbf{M}}_{t}^{1} - \mbox{\textbf{M}}_{t-1}^{1}\right),
\left( \mbox{\textbf{M}}_{t}^{2} - \mbox{\textbf{M}}_{t-1}^{2}  \right), \ldots,
\left( \mbox{\textbf{M}}_{t}^{l} - \mbox{\textbf{M}}_{t-1}^{l}  \right) 
\right\}
\end{equation}
where $\small\mbox{\textbf{M}}_{t}^{l}$ represents the feature maps of $\small{\mbox{\textbf{X}}}_{t}^{a}$ that are extracted from $\textit{l}$ layer in the spatial stream network. Hence, the output of the temporal sub-network is given by:
\begin{equation}
\footnotesize
{\textbf{\textit{Pred}}_{temporal} =  \frac{1}{N-1}\sum_{t=2}^{N}\mathcal{P_{T}} \left( {\textbf{\mbox{X}}}_{t}^{d}, {\mathcal{D}\left(  \widetilde{\mbox{\textbf{M}}}_{t}, \widetilde{{\mbox{\textbf{M}}}}_{t+1}\right)}, {{\textbf{W}}}_{T}\right)}.
\end{equation}
where \YZ{$\mathcal{P_{T}}\left(\cdot\right)$ denotes the mathematical expression of the temporal network with parameters $\small \mbox{\textbf{W}}_{T}$}.

\subsubsection{Optimization and Coupled Spatio-temporal Feature} We train the STN on the melanoma diagnosis task with all images of each sequential set. We apply a sigmoid function to the averaged output of the \YZ{STN} and optimize the entire model with binary cross-entropy loss. Therefore, the proposed \YZ{STN} can track the dermoscopic changes over time using clues provided by the temporal difference information from both the raw pixels and the abstract features. Once the model is well-trained, we construct a series of coupled spatio-temporal features $\boldsymbol{F} = \left \{ f_{{ST}_{1}}, f_{{ST}_{2}}, ..., f_{{ST}_{N}} \right \} \in \mathbb{R}^{N \times C}$ by concatenating \YZ{output} from the penultimate layers of the two subnetworks for the subsequent early diagnosis task. To elaborate, at time $t$, the feature $f_{{ST}_{t}} $ contains spatial appearance characterisation from the current lesion image and abstraction of temporal changes associated with the previous lesion image.

\subsection{Early Diagnosis Module}
The early diagnosis module evaluates spatio-temproal features from follow-up images of lesions and provides predictions at individual time points as the lesion progresses. This is entirely different setting compared with all previous algorithms for early melanoma detection which are developed on static images.

An early diagnosis model should be capable of accurately predicting a lesion's category at any given time point from a series of inputs. However, achieving this can be difficult for several reasons: 1) prediction accuracy for early predictions inevitably tends to be worse than that of later stages because of insufficient clues regarding lesion evolution; 2) inconsistent prediction scores from various time points as models can be easily disturbed by noise (e.g., lesion misalignments and lighting differences) and by its own model uncertainty~\cite{gal2016dropout}.    
Therefore, we propose to deal with these issues by designing a sequential context aggregation block based on an intra-attention mechanism, as well as a customized training strategy using temporal knowledge distillation.

\subsubsection{Sequential Context Aggregation Block} In practice, evaluating more images of a lesion can lead to a more consistent and confident diagnostic result. This means that the confidence score of a lesion should be monotonic for either melanoma or benign lesions, that is, benign lesions' scores should continue to decrease or remain unchanged, whereas malignant melanoma should have increased prediction scores. To maintain this consistency, it is crucial to correlate the relationship of the features across time and regulate them in an adaptive way for decision making.

\begin{figure}[ht]
\centering
\includegraphics[width=0.48\textwidth]{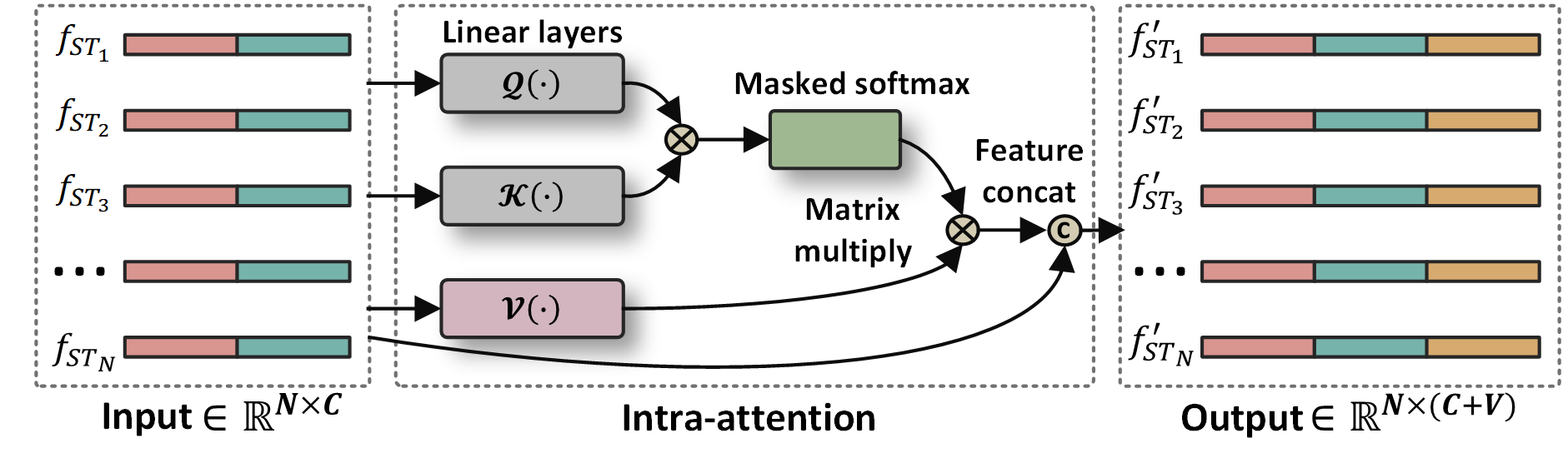}
\caption{Detail of the sequential context aggregation block.}
\label{fig: attention}
\end{figure}

Thus, we propose to aggregate features from different time points with a masked sequential context aggregation block (SCA) \YZ{which is mainly inspired by \cite{mishra2018simple}}. The SCA block consists of three linear transformation layers and a masked softmax layer. Similar to \cite{vaswani2017attention}, for a serial input of spatio-temporal features $\small{\textit{\textbf{F}} = \left \{ \textbf{\textit{f}}_{ST_{1}},  \textbf{\textit{f}}_{ST_{2}}, ..., \textbf{\textit{f}}_{ST_{N}} \right \}\in \mathbb{R}^{N \times C}}$, we first generate weights $\textbf{\textit{A}}= \left \{ \textbf{\textit{a}}_{1}, \textbf{\textit{a}}_{2},..., \textbf{\textit{a}}_{N} \right \}\in \mathbb{R}^{N}$ by applying a masked softmax on the matrix of query-key mapping:
\begin{equation}
\footnotesize
\textbf{\textit{A}}= \textit{\textbf{Softmax}}\left(\frac{\mathcal{Q} \left( \textit{\textbf{W}}_{\textit{q}}, \textit{\textbf{F}} \right) \cdot \mathcal{K} \left( \textit{\textbf{W}}_{\textit{k}}, \textit{\textbf{F}} \right)^{T}}{\sqrt{K}}\cdot \textit{\textbf{Mask}}\right)
\end{equation}
\begin{equation}
\footnotesize
\YZ{
\textit{\textbf{Mask}} = \begin{pmatrix}
 1&  -inf&  ...& -inf &-inf\\ 
 1& 1 & ... & -inf &-inf\\ 
 ... &  ... &  ...& ... & -inf\\ 
 1&  1&  ...& ...&1
\end{pmatrix}
}
\end{equation}
Then, we compute the weighted features and further concatenate them with the original input features:
\begin{equation}
\footnotesize
\textit{\textbf{F}}' = \textbf{\textit{Concat}}\left (\textit{\textbf{F}}, \textit{\textbf{A}}\cdot \mathcal{V}\left ( \textit{\textbf{W}}_{v},\textit{\textbf{F}} \right )\right)
\end{equation}

where $\textbf{\textit{Mask}} \in \mathbb{R}^{N \times N}$ is used to zero out future values so that a certain time point's query vector cannot make use of unseen feature information; $\small{\mathcal{Q}\left(\cdot \right )}$, $\small{\mathcal{K}\left(\cdot \right )}$, and $\small{\mathcal{V}\left(\cdot \right )}$ denote linear transformation functions, and $\footnotesize {\textit{\textbf{W}}_{\textit{q}}}\in \mathbb{R}^{N\times K}$, $\footnotesize{\textit{\textbf{W}}_{\textit{k}}} \in \mathbb{R}^{N\times K}$ and $\footnotesize{\textit{\textbf{W}}_{\textit{v}}} \in \mathbb{R}^{N \times V}$are the corresponding parameters; In our study, we set $\textit{K}=\textit{V}=16$, hence the final dimension of each output feature in $\small{\textit{\textbf{F}}' = \left \{ \textbf{\textit{f}}^{'}_{ST_{1}}, \textbf{\textit{f}}^{'}_{ST_{2}}, ..., \textbf{\textit{f}}^{'}_{ST_{N}} \right \}\in \mathbb{R}^{N \times \left(C+V\right)}}$ is 48.

\subsubsection{Temporal Knowledge Distillation} To reduce the gap in prediction accuracy between early and late predictions, we propose a knowledge distillation based training strategy to distill tendency knowledge from a later time point to an earlier time point. During training, we incorporate a constraint term into the objective function to penalize the dissimilarity between the predictions at different time points:  
\begin{equation}
\footnotesize
\mathcal{L} = \alpha \cdot \mathcal{L}_{tkd} + \left ( 1-\alpha \right ) \cdot \mathcal{L}_{bce}
\end{equation}
\begin{equation}
\footnotesize
\begin{aligned}
\mathcal{L}_{tkd} = -\frac{1}{M}\sum_{i=1}^{M}\sum_{t=1}^{N-1} \left ( \tilde{y}_{N} \cdot \log \delta \left ( \textbf{\textit{w}}_{clss} \cdot \textbf{\textit{f}}'_{ST_{t}} \right)\right ) + \\ \left( 1 - \tilde{y}_{N} \right ) \cdot \log \left( 1-\delta \left ( \textbf{\textit{w}}_{clss} \cdot \textbf{\textit{f}}'_{ST_{t}} \right) \right ) - \mathcal{L}^{\ast}_{\tilde{y}_{N}}
\end{aligned}
\end{equation}
\begin{equation}
\footnotesize
\begin{aligned}
\YZ{\mathcal{L}_{bce} = -\frac{1}{M}\sum_{i=1}^{M}\sum_{t=1}^{N} \left ( y_{i} \cdot \log \delta \left ( \textbf{\textit{w}}_{clss} \cdot \textbf{\textit{f}}'_{ST_{t}} \right)\right ) +} \\ \YZ{\left( 1 - y_{i} \right ) \cdot \log \left( 1-\delta \left ( \textbf{\textit{w}}_{clss} \cdot \textbf{\textit{f}}'_{ST_{t}} \right) \right )}
\end{aligned}
\end{equation}
where $\alpha$ is the coefficient between the distillation loss $\mathcal{L}_{tkd}$ and binary cross-entropy loss $\mathcal{L}_{bce}$;
$\small{\mathcal{L}^{\ast}_{\tilde{y}_{N}}}$ is the entropy of predictions from the last time point which only serves the purpose of simplifying notation and will not affect the optimisation; \YZ{$y_{i}$ is the ground truth label of the \textit{i}-th image sequence}; and $\textbf{\textit{w}}_{clss}$ denotes the weights of the final classifier which shares the same values across time points.

\subsubsection{Prediction and output mechanism} During training stage, we fix the length of the input image sequences by padding the initial screening image or by randomly sampling the required number of images. We optimize the early diagnosis module by jointly minimizing the disagreement between the ground-truth and the predictions from different time points. Once the module is well-trained, we compute the decision thresholds at each time point according to the maximum Youden’s index (sensitivity + sensitivity - 1) \cite{fluss2005estimation}. At the inference stage, as we sequentially input lesion images into our model, we need to generate prediction labels of a lesion over time to determine the transition point at which a benign lesion evolves into melanoma. We achieve this by designing an output mechanism that cumulatively compares the probability scores and the thresholds at consecutive time points. For image sequences having a length larger than the input length of our model, we generate a series of fixed-length overlapped sub-sequences and then vote on their prediction labels to obtain the prediction labels of the entire sequence.

\section{Experiments and Results}
\subsection{Dataset and Implementation}
\subsubsection{Dataset and evaluation}In this study, we collected 179 serial dermoscopic imaging data from 122 patients, including a total of 730 dermoscopic images. The dataset is well-balanced and consists of 90 benign lesions and 89 malignant lesions (including both invasive and in situ melanoma). Each lesion undergoing digital dermoscopic imaging monitoring was eventually excised due to clinical concerns and subsequently verified by pathological examination. The length of the dermoscopic image sequences varied from 1 to 12, and the average number of images in each image sequence was approximately 4.

We performed five-fold cross-validation to evaluate our method. Specifically, we first randomly partitioned the entire dataset into five folds. During each round of cross-validation, we selected one fold as the testing set and further split the remaining part of the data into the training and validation sets (90\% for training and 10\% for validation). The testing set was successively selected from the five-fold data, which means that each individual lesion would be used for testing after the five-fold cross-validation. In addition, the hyper-parameters and models were only trained with the training set and the validation set, that is, the testing set was never utilized to select a model. Training details are provided in the Appendix (Section A).

\subsubsection{Baseline models for comparison} \YZ{We implemented four deep learning-based baselines for performance comparison: 1) \textbf{Single-img-CNN:} The Single-img-CNN was trained with lesion images of single time without considering the temporal information; 2) \textbf{CNN-Score-Fusion:} The CNN-Score-Fusion model is similar to the Single-img-CNN, except during test phase we incorporated temporal clues by averaging the disease prediction scores of images within the input sequence; 3) \textbf{CNN-Feature-Pooling:} The CNN-Feature-Pooling model was directly trained with sequential images by combining the CNN features of individual images via average pooling; 4) \textbf{CNN-LSTM:} The CNN-LSTM model, trained on the image sequence, performed temporal aggregation over the CNN features of sequential dermoscopic images using LSTM. All models used the same ImageNet pre-trained ResNet-34 as the backbone. We provide the details of these models in the Appendix.}

\subsubsection{Interaction platform for reviewers} We invited 12 clinicians to evaluate our serial dermoscopic data and the diagnostic results were compared to that of our model. The serial images were displayed to the reviewers using Qualtrics™ (Provo, UT, USA). The reviewers were blinded to the patient diagnoses. Information provided for each case included age, sex, location of the lesion, and date of imaging. The reviewers were initially only shown the first dermoscopic image in the sequence and were asked to provide a diagnosis of either ‘benign’ or ‘malignant’. As the reviewers progressed through the sequence of images for each case, dermoscopic images were provided side-by-side to allow an assessment of changes. Prior responses could not be changed once the diagnosis was entered and submitted. Ten single time point melanoma images were included to reduce bias from reviewers, in which they might assume the first serial image in any case series to be benign. 

\begin{figure*}[ht]
\centering
\includegraphics[width=0.9\textwidth]{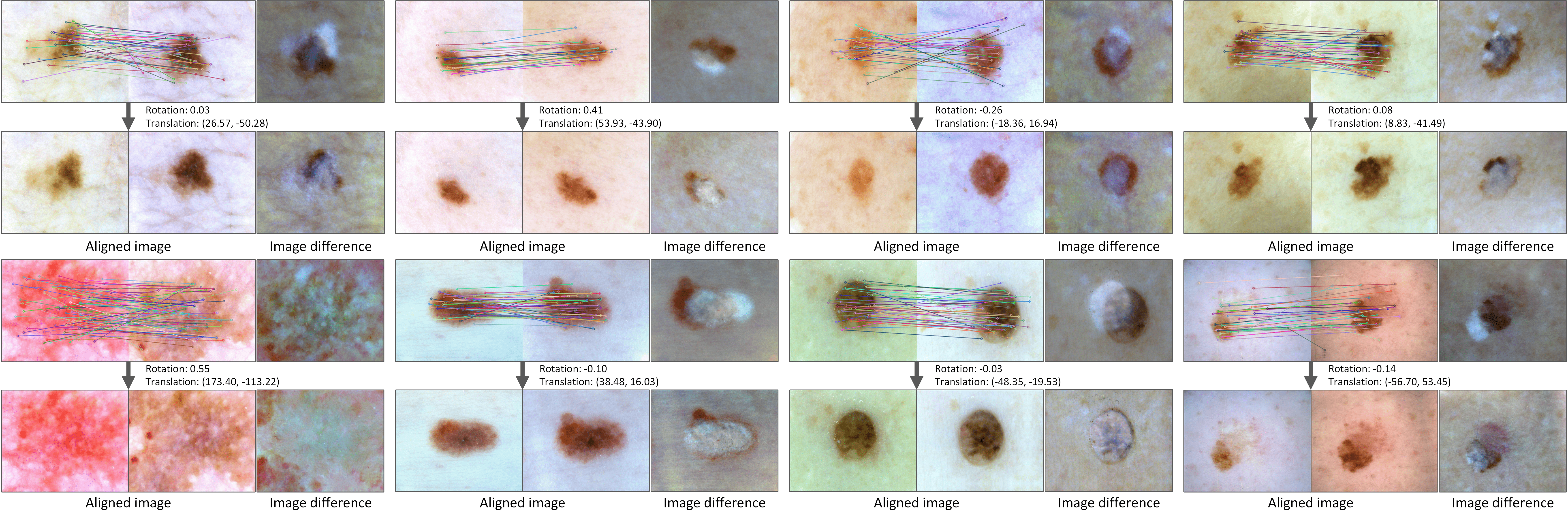}
\caption{ Results of dermoscopic image alignment. For each sample, we calculated the estimated transformation parameters, including the rotation and translation. We also show the image differences listed in the rightmost column without alignment. The pixel value in white indicates the mismatch region between the reference and moving images.}
\label{fig: alignment}
\end{figure*}

\begin{table*}[ht]
\centering
\caption{Results of the comparison study and ablation studies, reported on image sequences with a length of 4.}
\begin{tabular}{|c|c|c|c|c|c|}
\hline 
\footnotesize
Methods & Accuracy(\%) & AUC(\%) & Precision(\%) & Sensitivity(\%) & Specificity(\%) \\
\hline \hline
Single-img-CNN & 61.24$\pm$6.54 & 66.76$\pm$9.87 & 61.51$\pm$8.16 & 58.53$\pm$6.59 & 63.79$\pm$6.39\\
\hline 
CNN-Score-Fusion & 60.67$\pm$8.04 & 67.13$\pm$8.31 & 60.85$\pm$8.29 & 60.02$\pm$ 7.95& 61.53$\pm$8.34\\
CNN-Feature-Pooling & 60.67$\pm$7.05& 66.09$\pm$6.62 & 61.12$\pm$9.39 & 57.21$\pm$6.77 & 63.87$\pm$7.80\\
CNN-LSTM & 64.47$\pm$11.91 & 68.69$\pm$11.78 & 65.76$\pm$11.87 & 62.31$\pm$10.60 & 66.85$\pm$12.69\\
\hline
\multicolumn{6}{|l|}{Proposed model without alignment} \\ \hline
\YZ{Only spatial network} & 62.79$\pm$7.42 & 65.41$\pm$6.72 & 63.48$\pm$14.86 & 57.66$\pm$8.53 & 67.86$\pm$7.15 \\
\YZ{Only temporal network} & 62.79$\pm$4.81 & 68.72$\pm$7.75 & 63.00$\pm$11.35 & 60.52$\pm$4.64 & 65.48$\pm$5.22 \\
\YZ{Spatio-temporal network} & 67.19$\pm$6.05 & 70.06$\pm$7.53 & 67.03$\pm$14.37 & 63.93$\pm$6.89 & 70.42$\pm$6.34 \\
\YZ{Spatio-temporal network with interconnected setting} & 67.11$\pm$11.01 & 73.25$\pm$14.12 & 68.30$\pm$11.92 & 63.79$\pm$10.51 & 70.42$\pm$11.53 \\ \hline
\multicolumn{6}{|l|}{Proposed model with alignment} \\ \hline
\YZ{Only spatial network} & 62.89$\pm$5.78 & 68.98$\pm$6.85 & 63.20$\pm$10.47 & 60.45$\pm$5.76 & 65.64$\pm$6.25\\
\YZ{Only temporal network} & 65.54$\pm$7.14 & 70.63$\pm$11.16 & 66.72$\pm$9.51 & 61.85$\pm$6.03 & 69.20$\pm$8.43\\
\YZ{Spatio-temporal network} & 68.11$\pm$7.21 & 71.73$\pm$6.21 & 65.89$\pm$12.07 & 66.89$\pm$10.43 & 68.30$\pm$4.52\\
\YZ{Spatio-temporal network with interconnected setting} & \bfseries{69.98$\pm$10.48} & \bfseries{74.34$\pm$10.83} & \bfseries{71.61$\pm$13.53} & \bfseries{69.66$\pm$9.68} & \bfseries{70.99$\pm$12.47}\\
\hline
\end{tabular}\label{tab: seq_diag_cmp}
\end{table*}

\begin{figure*}[ht]
\centering
\includegraphics[width=0.9\textwidth]{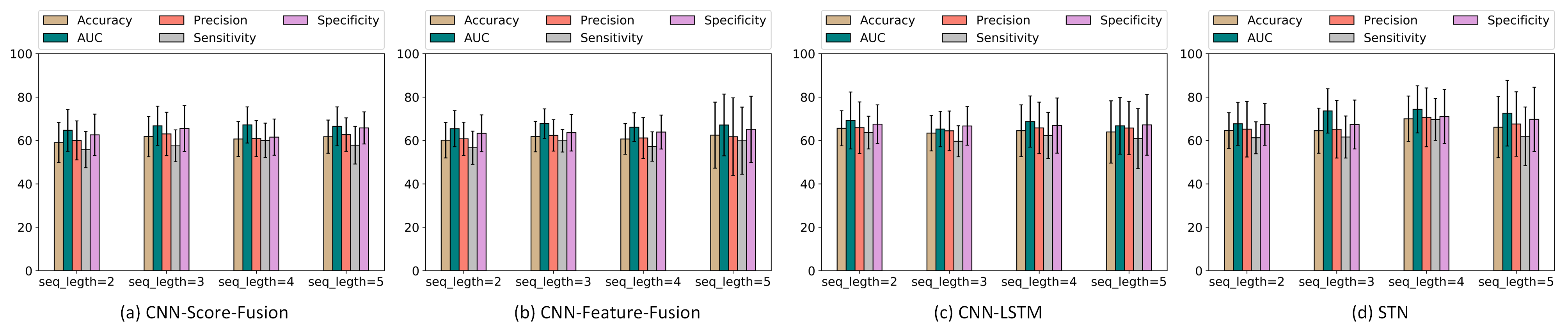}
\caption{Comparison results of the sequence learning models when varying the length of the training image sequence.}
\label{fig: diag_seq_auc}
\end{figure*}

\subsection{Results of Skin Lesion Alignment}
\YZ{Lesion alignment pre-processes the image for the subsequent extraction of the lesion growth region by computing image differences. We evaluated the performance of lesion alignment by providing aligned image pairs and corresponding image differences. Fig. \ref{fig: alignment} shows qualitative visual alignment results for consecutive dermoscopic images. We can see that the warped images and the reference images show strong location and content-wise consistency when compared to the unaligned samples. We then further verified the necessity of the alignment component by training the two-stream network for lesion diagnosis using aligned and unaligned sequential image sequences. The results are presented in Table \ref{tab: seq_diag_cmp}.}

\subsection{Result of Melanoma Diagnosis on Sequential Images}
\YZ{We evaluated the performance of the spatio-temporal network (STN) for melanoma diagnosis on sequential images. For each dermoscopic image sequence, we used the STN to predict melanoma using all images of each sequence. Our aim is to verify the effectiveness of the proposed STN and the benefit of incorporating temporal information in melanoma diagnosis. Notably, this task is different from the following early melanoma diagnosis task which needs to make predictions at each time point.}

\begin{table*}[ht]
\centering
\caption{The results of the comparison and ablation studies for early diagnosis. The AUC at each time point is reported.}
\begin{tabular}{|c|c|c|c|c|c|}
\hline
Features & \diagbox{Models}{AUC}{Times} & Time 1 & Time 2 & Time 3 & Time 4 \\ \hline \hline
\multirow{5}{*}{CNN spatial feature} & Single-img-CNN & 63.04$\pm$7.14 & 61.25$\pm$7.14 & 62.61$\pm$8.78 & 66.13$\pm$7.14 \\ \cline{2-6} 
                              & Single-img-CNN (HAM) & 65.47$\pm$7.46 & 64.22$\pm$6.77 & 65.54$\pm$7.14      & 70.26$\pm$ 3.35        \\ \cline{2-6} 
                              & CNN-Score-Fusion     & 60.59$\pm$9.88 & 65.07$\pm$4.19 &56.77$\pm$4.58 & 65.72$\pm$8.13\\ \cline{2-6} 
                              & CNN-Feature-Fusion   &   60.59$\pm$9.88   & 61.73$\pm$8.40 & 63.34$\pm$10.05 &  65.48$\pm$8.36      \\ \cline{2-6} 
                              & CNN-LSTM              & 69.79$\pm$9.64    & 69.13$\pm$8.29       & 68.29$\pm$15.36    &  65.92$\pm$6.99     \\ \hline
\multirow{4}{*}{Coupled spatio-temporal feature}  & CST-Baseline &  68.71$\pm$9.03     &  69.51$\pm$6.70   &  70.29$\pm$11.72   &  67.56$\pm$6.35   \\ \cline{2-6} 
                              & CST-LSTM  & 69.04$\pm$9.44 & 70.43$\pm$8.86 & 71.05$\pm$9.78 & 72.05$\pm$8.43   \\ \cline{2-6} 
                              & CST-SCA & 68.76$\pm$8.93 & 69.79$\pm$8.21 & 70.40$\pm$11.13 & 71.98$\pm$8.68
                              \\ \cline{2-6} 
                              & CST-SCA-TKD          &   70.11$\pm$8.22   &  70.81$\pm$7.90   &   71.75$\pm$10.17 & 72.73$\pm$7.70     \\ \hline
\end{tabular}\label{tab: early_diag}
\end{table*}

\begin{figure*}[ht]
\centering
\includegraphics[width=0.9\textwidth]{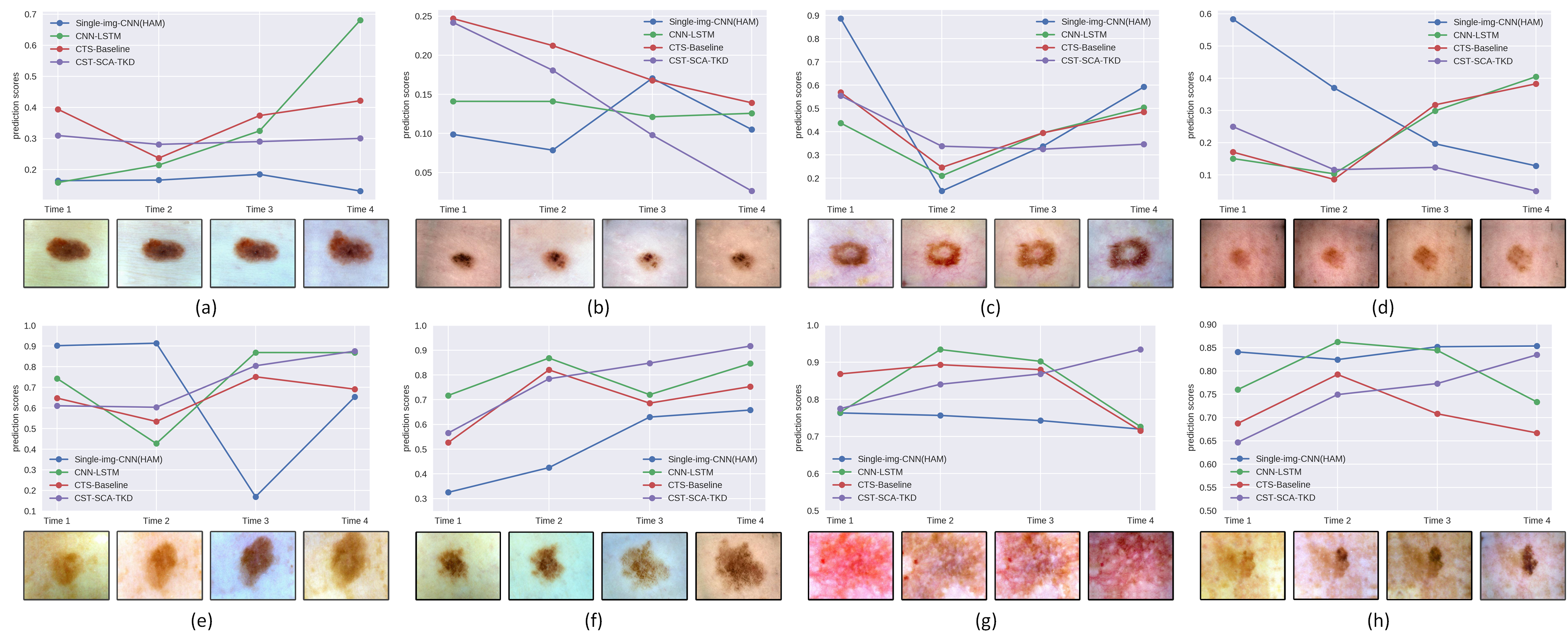}
\caption{Prediction scores of individual lesions from different models. (a)-(d) are benign lesions and (e)-(f) are malignant lesions.}
\label{fig: indivi_lesions}
\end{figure*}

\YZ{We first compare all the methods at the sequence length of $\textbf{N}$ = 4 in Table \ref{tab: seq_diag_cmp} which was the average length of our sequential dataset. Further results for various sequence lengths are shown in Fig.~\ref{fig: diag_seq_auc}. We found that all of the models using sequential images had better AUC than the Single-img-CNN trained with snapshot images, and the proposed model achieved the best performance with an accuracy of 69.98\%, AUC of 74.34\%, precision of 71.61\%, sensitivity of 69.66\%, and specificity of 70.99\%. The temporal stream network has better performance when compared to using just the spatial stream network. By combining the two-stream network, we obtained a significant performance improvement. Notably, by removing the interconnected setting in our model, the AUC was reduced by $\sim$1.7\%. Our model achieved a consistent AUC boost from 67.72\% to 74.34\% when increasing the sequence length from 2 to 4 \footnote{We equalized the required input length by adding the first screening image or randomly selected consecutive images.}. However, there was no obvious performance improvement for all the comparative sequential models when the sequence length was increased. These results demonstrate the benefit of incorporating temporal clues in melanoma diagnosis, and demonstrate the effectiveness of the proposed method in learning spatio-temporal features from serial images.}

\subsection{Evaluation of Early Melanoma Diagnosis}
We evaluated the effectiveness of each component designed for the early melanoma diagnosis task. We input the image sequence into our model, and compute probability of a lesion belonging to melanoma at each individual time point. We report the ablation results, as well as the comparative results with other models.

\begin{figure*}[ht]
\centering
\includegraphics[width=0.95\textwidth]{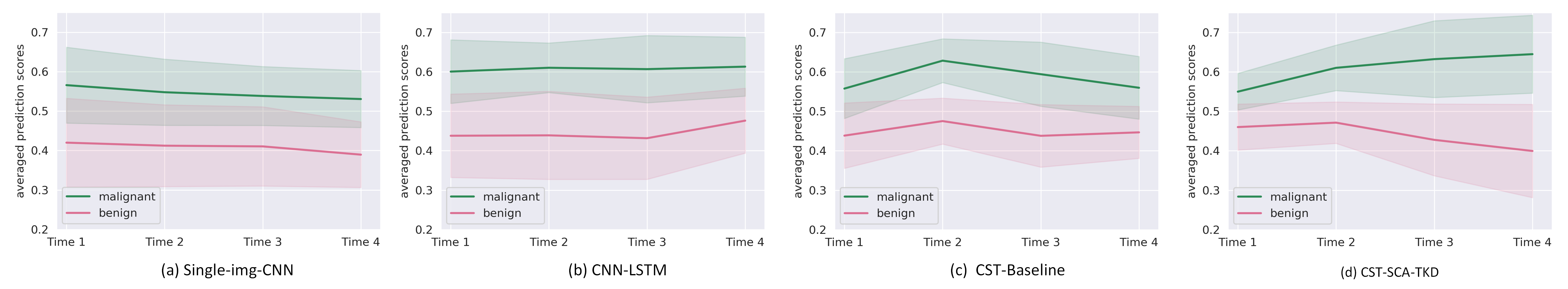}
\caption{Prediction scores of lesions from different models by varying time. We average the probability scores of all benign and malignant lesions, respectively. The proposed model shows clear trends for both classes.}
\label{fig: avg_predictions}
\end{figure*}

\begin{table*}[ht]
\centering
\caption{Comparison results of clinicians and our model.}
\begin{tabular}{|l|c|c|c|c|c|c|} \hline
\multirow{2}{*}{}                          & \multicolumn{3}{c|}{\textbf{Based on final image diagnosis}}      & \multicolumn{3}{c|}{\textbf{Based on first time malignant diagnosis}}                \\ \cline{2-7} 
                                           & Accuracy (\%)  & Sensitivity (\%)  & Specificity (\%)    & Accuracy (\%)   & Sensitivity (\%) & \multicolumn{1}{l|}{Specificity (\%)} \\ \hline
\multicolumn{7}{|l|}{Registrars (n=5) \YZ{; E$<$5} }                                                                                                                                            \\ \hline
Reviewer 1                                 & 54.75          & 85.39             & 24.44               & 55.87           & 87.64            & 24.44                                 \\ \hline
Reviewer 2                                 & 51.96          & 42.70             & 61.11               & 51.40           & 47.19            & 55.56                                 \\ \hline
Reviewer 4                                 & 56.42          & 48.31             & 64.44               & 58.10           & 59.55            & 56.67                                 \\ \hline
Reviewer 7                                 & 43.58          & 59.55             & 27.78               & 43.58           & 62.92            & 24.44                                 \\ \hline
Reviewer 12                                & 51.96          & 75.28             & 28.89               & 52.51           & 82.02            & 23.33                                 \\ \hline
\multicolumn{7}{|l|}{Dermatologists (n=7) \YZ{; E$\geq$5} }                                                                                                                                        \\ \hline
Reviewer 3                                 & 55.31          & 53.93             & 56.67               & 55.87           & 55.06            & 56.67                                 \\ \hline
Reviewer 5                                 & 51.96          & 71.91             & 32.22               & 51.96           & 74.16            & 30.00                                 \\ \hline
Reviewer 6                                 & 54.19          & 31.46             & 76.67               & 54.19           & 37.08            & 71.11                                 \\ \hline
Reviewer 8                                 & 54.19          & 79.78             & 28.89               & 53.07           & 80.90            & 25.56                                 \\ \hline
Reviewer 9                                 & 56.98          & 67.42             & 46.67               & 56.98           & 69.66            & 44.44                                 \\ \hline
Reviewer 10                                & 59.78          & 64.04             & 55.56               & 57.54           & 68.54            & 46.67                                 \\ \hline
Reviewer 11                                & 60.89          & 64.04             & 57.78               & 60.34           & 65.17            & 55.56                                 \\ \hline
\multicolumn{7}{|l|}{Average results}                                                                                                                                              \\ \hline
Registrars (n=5)                           & 51.73          & 62.25             & 41.33               & 52.29           & 67.87            & 36.89                                 \\ \hline
Dermatologists (n=7)                       & 56.19          & 61.80             & 50.63               & 55.71           & 64.37            & 47.14                                 \\ \hline
All (n=12)                                 & 54.33          & 61.99             & 46.76               & 54.28           & 65.82            & 42.87                                 \\ \hline
Our model                                  & 63.69          & 60.67             & 66.67               & 61.45           & 75.28            & 47.78                                 \\ \hline
\end{tabular}\label{tab: human_ai_cmp}
\end{table*} 

\begin{figure}[ht]
\centering
\includegraphics[width=0.48\textwidth]{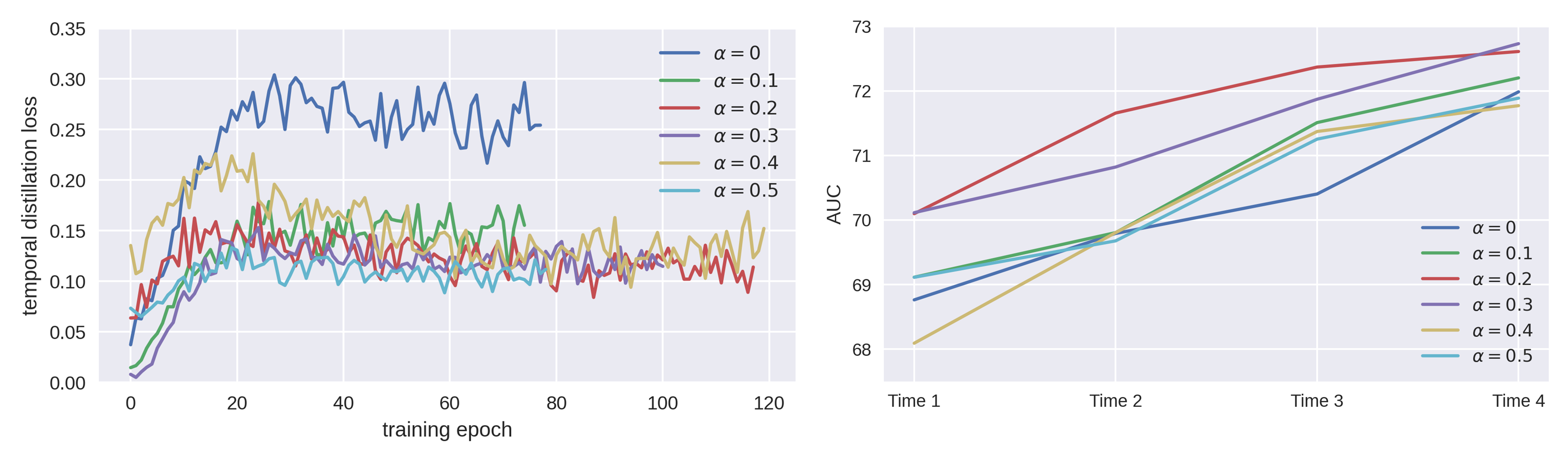}
\caption{Loss and AUC of our model by varying the coefficient of the proposed temporal distillation loss.} 
\label{fig: ablation_tkd}
\end{figure}

\noindent\textbf{Results:} Apart from the four baseline models, we also pre-trained a model with HAM-10000 \cite{tschandl2018ham10000}, and then further fine-tuned it on our sequential dermoscopic data. We refer to this model as Single-img-CNN(HAM). Regarding the proposed coupled spatio-temporal feature (CST), CST-Baseline denotes directly learning a classifier at each time point with the CST features of that time. CST-LSTM refers to the computation of the probability score at individual time points using LSTM. CST-SCA-TKD and CST-SCA are the proposed early diagnosis modules trained with and without the temporal knowledge distillation strategy, respectively. The results are provided in Table~\ref{tab: early_diag}, which shows that the performance of all the baseline models was inferior to that of the proposed model. In addition, we observed that the AUC of these models when taking more time points of images, that is, when incorporating more clues regarding lesion growth, did not show consistent improvement. For example, the CST-Baseline showed an increase in performance before Time 3, but the AUC decreased at Time 4. The AUC at Time 4 was worse than that at Time 1 by $\sim$1.2\%.
When large-scale external data were used, the single-img-CNN (HAM) achieved significant improvements of $\sim$2\%, $\sim$3\%, $\sim$3\%, and $\sim$4\% at the four different time points, respectively, compared with that of Single-img-CNN. However, the performance was still inferior to our method, and the AUC was even worse than our CST-baseline model, which demonstrates the necessity of including CST features for this task. 

\noindent\textbf{Prediction Consistency with the Aggregation Mechanism:} To visualize the prediction score trend for each image sequence, we present the prediction results of each lesion in Fig. \ref{fig: indivi_lesions}. We can observe that the prediction scores of CST-Baseline show inconsistent fluctuations across various time points. We speculate that the reason for this is that some lesions do not change evenly over time, and thus lesion evolution among consecutive images does not always show consistent, linear changes. In this case, the lesion growth captured by our CST features will vary according to their discriminability. In contrast, the proposed CST-SCA obtained a consistent performance improvement over time with an AUC of 68\% at Time 1, increasing to an AUC of 71.98\% at Time 4. In Fig. \ref{fig: avg_predictions}, we visualize the average prediction scores of benign lesions and malignant melanoma in the test data, respectively. The overall predictions of benign lesions remained unchanged and gradually decreased over time, whereas the predictions of malignant lesions gradually increased over time. This result verifies the effectiveness of the proposed aggregation mechanism in tracking lesion evolution and maintaining prediction consistency.

\noindent\textbf{Temporal Knowledge Distillation:} We evaluated the influence of incorporating the temporal knowledge distillation in our model. As listed in Table~\ref{tab: early_diag}, the CST-SCA-TKD obtained AUCs of 70.11\%, 70.81\%, 71.75\%, and 72.73\% from Time 1 to Time 4, respectively. Compared with that of CST-SCA, the TKD training strategy provided an improvement of $\sim$1.4\%, $\sim$1.1\%, $\sim$1.3\%, and $\sim$0.8\% at each of the four time points. Moreover, the AUC gap between Time 1 and Time 4 decreased from $\sim$3.2\% to $\sim$2.6\%. Fig.~\ref{fig: ablation_tkd} shows the effect of the coefficient $\alpha$ on the objective function (Eq. (8)). By increasing $\alpha$ from 0.1 to 0.5, the performance of early predictions first increases and then slightly decreases, and $\alpha=0.2$ gives the best performance. In Fig. \ref{fig: ablation_tkd}, we plot the training loss results under different $\alpha$ settings, and we can observe that TKD significantly reduces the divergence between early predictions and later predictions. 

\begin{figure*}[ht]
\centering
\includegraphics[width=0.98\textwidth]{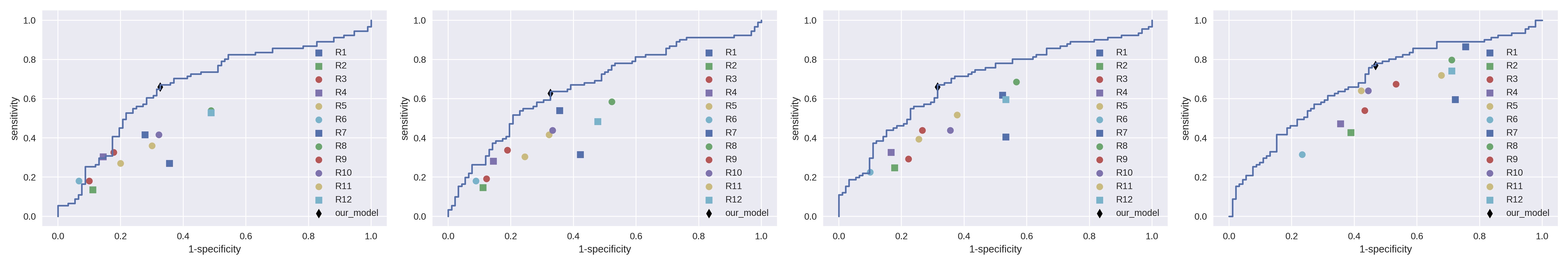}
\caption{Comparison of diagnostic results across time and later time points indicate that reviewers and the model are presented with more images of a lesion. Figures from left to right show the results at Time 1 to Time 4, respectively. Both human reviewers and our model tended to perform better when accessing more information regarding lesion evolution.}
\label{fig: roc_time}
\end{figure*}

\begin{figure*}[ht]
\centering
\includegraphics[width=\textwidth]{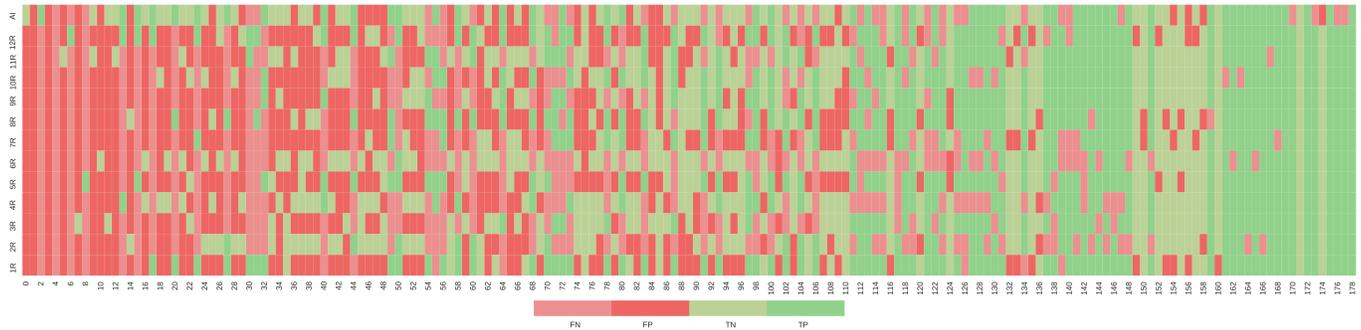}
\caption{Lesions arranged by the level of difficulty determined by the number of clinicians who correctly identified the case.}
\label{fig: diffcult}
\end{figure*}

\begin{figure}[ht]
\centering
\includegraphics[width=0.48\textwidth]{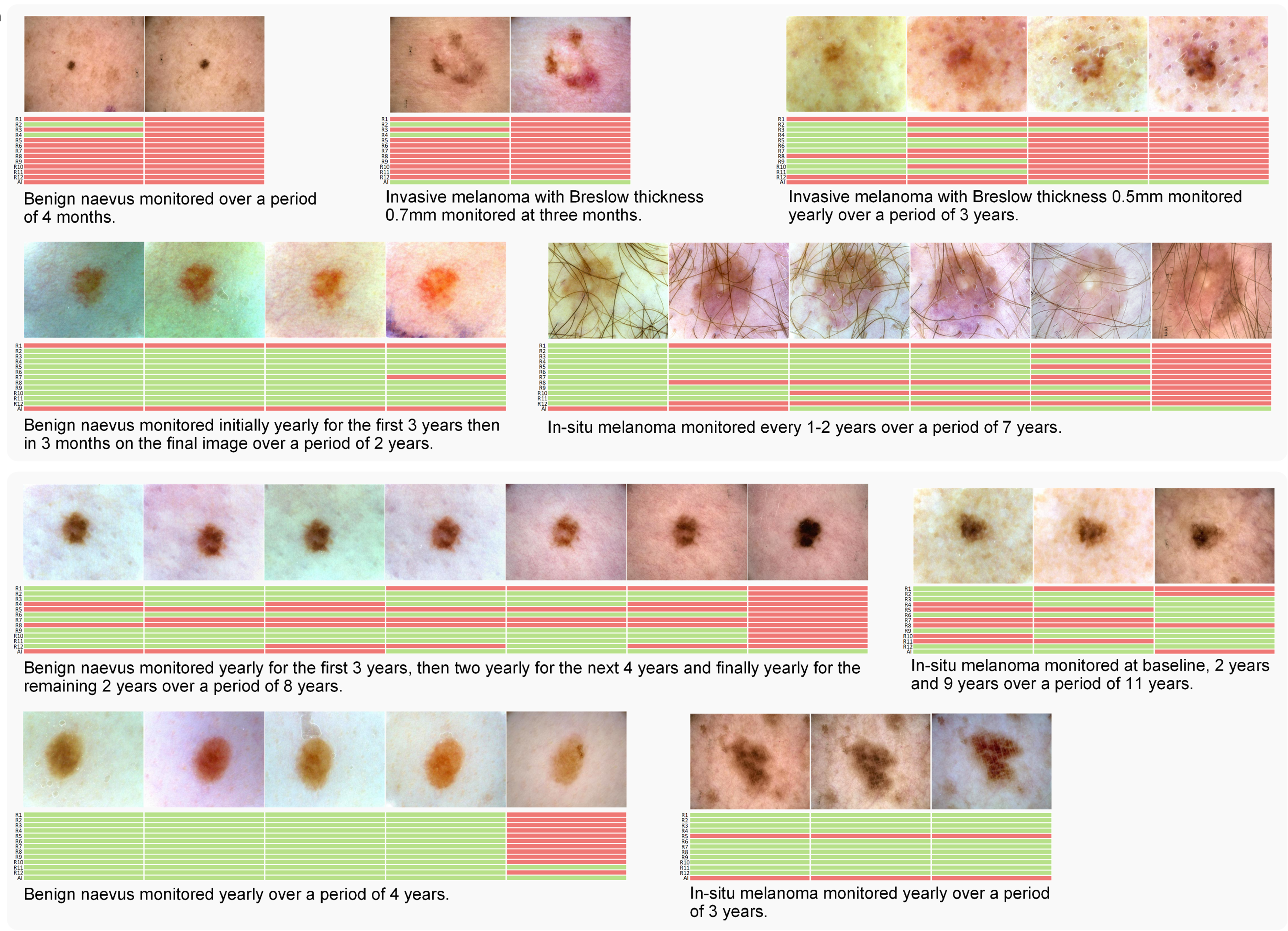}
\caption{The top five image sequences show lesions that were correctly but incorrectly diagnosed by the clinicians and our model. The four image sequences below show lesions that were incorrectly diagnosed by clinicians but correctly diagnosed by our model. Red and green bars below each lesion represent malignant and benign predictions for individual images, respectively.} 
\label{fig: pred_indivi}
\end{figure}

\begin{figure*}[ht]
\centering
\includegraphics[width=\textwidth]{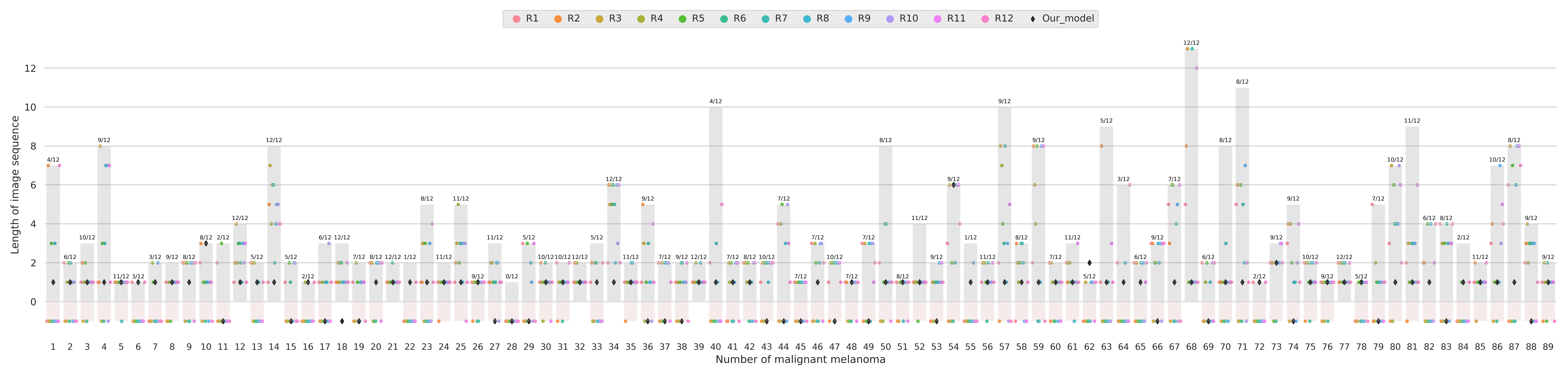}
\caption{Early diagnosis results for all melanomas for the 12 clinician reviewers (coloured dots) compared to the proposed model(black diamonds). The horizontal axis denotes the lesion ID and the vertical axis represents the length of the image sequence. Dots are placed at the image number in the sequence corresponding to the time point at which the correct diagnosis was made. The pink bar denotes a failure to make a correct diagnosis of melanoma \YZ{(best viewed while zoomed in)}.}
\label{fig: mel_diag_date}
\end{figure*}

\begin{figure*}[ht]
\centering
\includegraphics[width=\textwidth]{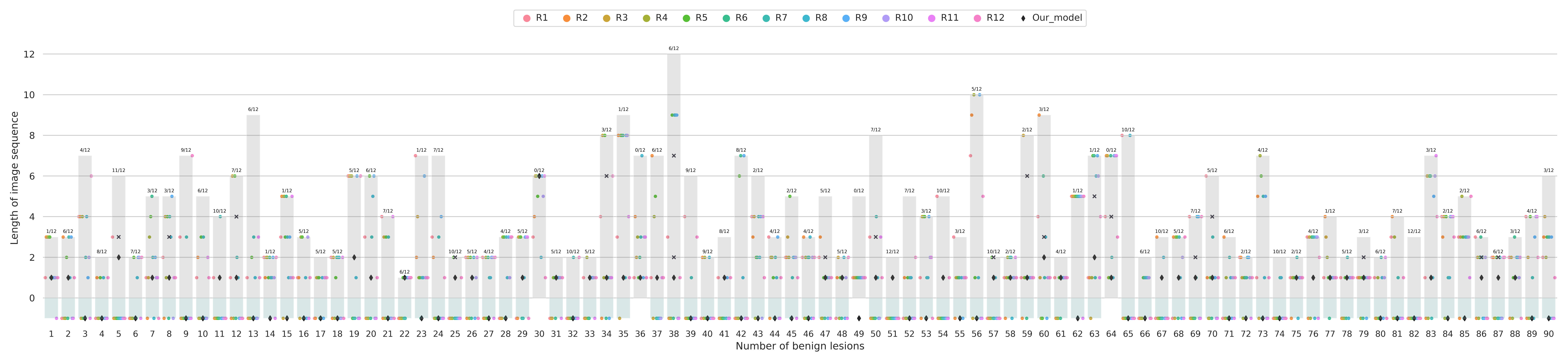}
\caption{Early diagnosis results for all benign lesions for the 12 clinician reviewers (coloured dots) compared to the proposed model(black diamonds). The horizontal axis denotes the lesion ID and the vertical axis represents the length of the image sequence. Dots are placed at the image number in the sequence corresponding to the time point at which the melanoma diagnosis was made. Cross marks indicate the image number wherein the AI model gave a diagnosis that changed from melanoma to a benign lesion. The green bar denotes the correct diagnosis of benign lesions \YZ{(best viewed while zoomed in)}.}
\label{fig: ben_diag_date}
\end{figure*}

\subsection{Compared with human results} 
In this section, we compare the early diagnosis performance of our model with that of human reviewers. The comparison includes the overall diagnostic accuracy and the time point at which the malignant lesions were correctly diagnosed. The serial dermoscopic image dataset was reviewed by 12 reviewers, including seven experienced dermatologists and five registrars from the Victorian Melanoma Service \footnote{\YZ{The five registrars’ experience was less than five years ($E<5$), whereas the dermatologists each had more than five years of experience ($E\geqslant5$).}}.

As shown in Table \ref{tab: human_ai_cmp}, based on the diagnosis made on the final image of each sequential set, clinicians achieved an overall accuracy, sensitivity, and specificity of 54.33\%, 61.99\%, and 46.76\%, respectively. Consultant dermatologists were more accurate and had better accuracy and specificity than registrars by 4.45\% and 9.30\%, respectively, although both had similar sensitivities. Compared to human reviewers, our model performed better with respect to accuracy, demonstrating an accuracy of 63.69\% which was 9.36\% higher than the clinician’s overall accuracy and 2.79\% greater than that of the one clinician with the highest accuracy. Notably, the algorithm exhibited a specificity of 66.67\% which was 19.82\% higher than that of the clinicians. The sensitivity of our model was similar to that of clinicians at 60.67

\YZ{Additionally, we reviewed the diagnostic accuracy for invasive and in situ melanomas separately. Of the 89 melanomas, 34 (38.2\%) were invasive with a mean Breslow thickness of 0.5, and 55 (61.8\%) were in situ melanomas. Clinicians accurately diagnosed 67.9\% of the invasive melanomas (67.94\% for dermatologists and 68.23\% for registrars) and 58.3\% of the in situ melanomas (58.18\% for dermatologists and 58.54\% for registrars) compared to our model which correctly diagnosed 61.8\% of invasive melanomas and 60.0\% of in-situ melanomas. The median Breslow thickness of the invasive melanomas that were incorrectly diagnosed by the clinicians and the model was similar, at 0.3 mm.}

\YZ{Fig. \ref{fig: diffcult} shows the diagnostic result for each sequential set of lesions arranged by the level of difficulty determined by the number of clinicians who correctly identified the case. Our dataset contains a balance of ‘easy’ to diagnose lesions, which the majority of clinicians were able to diagnose correctly, and ‘difficult’ to diagnose lesions, which the majority of clinicians were unable to diagnose correctly. There was no clear correlation between the difficulty of the cases based on the clinicians’ responses and the correct diagnosis from the model. Of the 10 cases correctly diagnosed by all clinicians, four were incorrectly diagnosed using the algorithm. When examining these cases in detail together with the other cases, the model incorrectly diagnosed many smaller lesions and lesions with poorly defined borders}.

Although it is important to consider the results based on the final image diagnosis, in a real-world clinical setting, a malignant diagnosis would warrant a biopsy which would lead to the cessation of serial monitoring. Therefore, we also performed an analysis based on when a malignant diagnosis was first reported by either the clinicians or the algorithm. When comparing results from final image diagnosis to that of the first malignant diagnosis, clinicians had similar accuracy, and the model showed reduced accuracy from 63.69\% to 61.45\%. Both clinicians and the model had increased sensitivity and reduced specificity, but the reduction in specificity was more marked for the algorithm, as listed in Table \ref{tab: human_ai_cmp}. \YZ{Due to the dynamic changes in the serial lesions, both the clinicians and the algorithm altered their diagnoses over time with additional image information. These results suggest that lesions may develop abrupt changes that can result in a malignant diagnosis, but also may stabilize over time, leading clinicians and the algorithm to prefer a benign diagnosis. Examples of lesions correctly diagnosed by clinicians and incorrectly by the model (and vice versa) are shown in Fig. \ref{fig: pred_indivi}.}

To evaluate whether early melanoma recognition is possible, we also recorded the time point at which clinicians and the algorithm first made a malignant diagnosis in both melanoma and benign cases. As shown in Fig. \ref{fig: mel_diag_date}, our model frequently gave a diagnosis of melanoma at earlier time points compared to clinicians with 54 (60.7\%) melanomas detected by the algorithm on the first follow-up image, compared to 29 (32.7\%) by clinicians. However, this phenomenon was also observed in benign cases (Fig. \ref{fig: ben_diag_date}), with 42 (46.7\%) benign lesions incorrectly diagnosed as melanoma by the model on the first follow-up image. Of the 34 invasive melanomas, the algorithm was able to correctly identify 25 melanomas, of which 24 were detected on the first sequential image (mean Breslow thickness 0.5 mm), and one melanoma was detected on the second sequential image (Breslow thickness 0.3 mm). For lesions that were not correctly identified by the algorithm, the mean Breslow thickness was 0.5 mm). As shown in Fig. \ref{fig: roc_time}, both clinicians and the model performed better in cases containing a longer sequence of images.

\section{Discussion}
Sequential monitoring of melanocytic naevi is recommended for the monitoring of high-risk individuals to improve early detection and reduce unnecessary biopsies \cite{kittler2006identification, salerni2012benefits}. Here, we demonstrate a model which incorporates information from dynamic changes detected from sequentially monitored melanocytic lesions to facilitate the prediction and early diagnosis of melanoma.

\YZ{In previous studies of single time point melanocytic lesions, computer algorithms have demonstrated a sensitivity ranging from 82.0\% to 97.1\% and specificity ranging from 60.0\% to 78.8\% \cite{brinker2019deep, marchetti2018results, brinker2019, fink2020diagnostic}. In these previous studies, diagnostic performance was compared with dermatologists whose sensitivity ranged from 67.2\% to 90.6\% and specificity from 59.0\% to 71.0\% \cite{brinker2019deep, marchetti2018results, brinker2019, fink2020diagnostic}}. In our results, lower sensitivity and specificity values were expected compared to other studies of single time point melanocytic lesions \cite{brinker2019deep, marchetti2018results, brinker2019, fink2020diagnostic} because our dataset was more challenging, consisting of sequentially monitored lesions in high-risk individuals whose lesions were all ultimately excised. The image dataset included melanomas with subtle architectural changes to atypical benign lesions which displayed significant transformations, that is clinically atypical lesions that were confirmed as benign on histopathology, but which showed changes suspicious for melanoma over time. Despite this, our model’s performance was superior to that of experienced clinicians, at least under the test conditions. It is worth noting that all lesions were ultimately excised, and thus, the true sensitivity of clinicians is, in fact, much higher. The model’s superior specificity, however, suggests that unnecessary excision of benign lesions may be avoided in some cases.

Additionally, our model was able to detect melanoma earlier than clinicians, which provides proof of concept for the algorithm’s function as a prognostic tool to improve early detection. Based on this, in the future we aim to identify computer generated biomarkers that can categorize melanocytic lesions into high and low risk of evolving into melanoma over time. Low-risk lesions would not require ongoing monitoring, whereas high-risk lesions would require closer dermoscopic surveillance or excision. The algorithm could also play a useful role in providing an additional diagnostic opinion to augment that of clinicians.

\YZ{Despite the model’s overall superior diagnostic performance, clinicians were able to diagnose a higher proportion of invasive melanoma (67.9\%) compared to the model (61.8\%). Additionally, our model showed poorer accuracy for small lesions or those with undefined borders. A larger training dataset is likely to improve model performance, with greater exposure to these types of lesions. Currently, however, the use of the algorithm on a wide range of lesions with different morphologies is limited, which may in part explain why there was no clear correlation between the difficulty of cases determined by the clinician’s responses and the accuracy of the model. Although the algorithm was able to correctly identify some ‘difficult’ lesions, the inverse was also true. It is important to understand which lesions may be misclassified by an algorithm so that clinicians are not led astray by the algorithm. There is a significant body of work to improve transparency of algorithms for precisely this reason \cite{tschandl2020human}.}

\YZ{There are some limitations in the interpretation of our study’s results. First, the clinicians’ assessment would not have been reflective of the real-life clinical setting. Clinicians in this study reviewed the images in an artificial environment without the context of the individual’s broader naevus ecosystem. Other valuable information related to melanoma risk (e.g., family history, past history, other phenotypic features) may impact diagnostic accuracy and the decision to excise a lesion \cite{haenssle2016association, haenssle2020man}. Additionally, our study included a balanced dataset of benign and malignant lesions which is useful in the training and evaluation of the algorithm; however, in the real clinical world, a very small percentage of the monitored lesions will be melanoma. Therefore, we acknowledge that a dataset enriched for malignancy, such as ours, will artificially enhance accuracy compared to real-world performance. Further validation of our study’s results with a larger prospective dataset as well as evaluation of the algorithm’s performance in a clinical setting is necessary. Regardless, we have demonstrated novel methods to train deep neural networks to monitor high-risk lesions with results that could revolutionise the approach to melanoma surveillance and screening.}

\section{Conclusion}
In this study, we present a framework for early melanoma diagnosis by modelling lesion growth using sequential dermoscopic images. We demonstrate the benefit of incorporating temporal clues in melanoma diagnosis, and demonstrate the superiority of the proposed method in capturing lesion changes from serial images for early melanoma detection, compared to other sequence models. In addition, we compare diagnostic performance of our algorithm to that of 12 clinicians. The result suggests that the algorithm is capable of consistently identifying melanoma at a clinicians’ standard, without the risk of over-reporting benign lesions. Additionally, the proposed model can predict melanoma earlier than can clinicians, which provides a proof of concept for the algorithm’s function as both a diagnostic and prognostic tool. Our approach has the potential to assist clinicians in more effective dermoscopic monitoring of high-risk patients. The benefits include reducing excessive screening by discontinuing sequential monitoring of benign lesions with a low probability of malignant transformation and potentially aiding more timely excision of lesions prior to an invasive malignant process.

\bibliographystyle{IEEEtran}
\bibliography{IEEEabrv, reference.bib}

\newpage
\clearpage
\newpage
\section{Appendix}
\subsection{Configuration of Models and Training Details}
\subsubsection{Configuration of models} 
\begin{figure}[ht]
\centering
\includegraphics[width=0.48\textwidth]{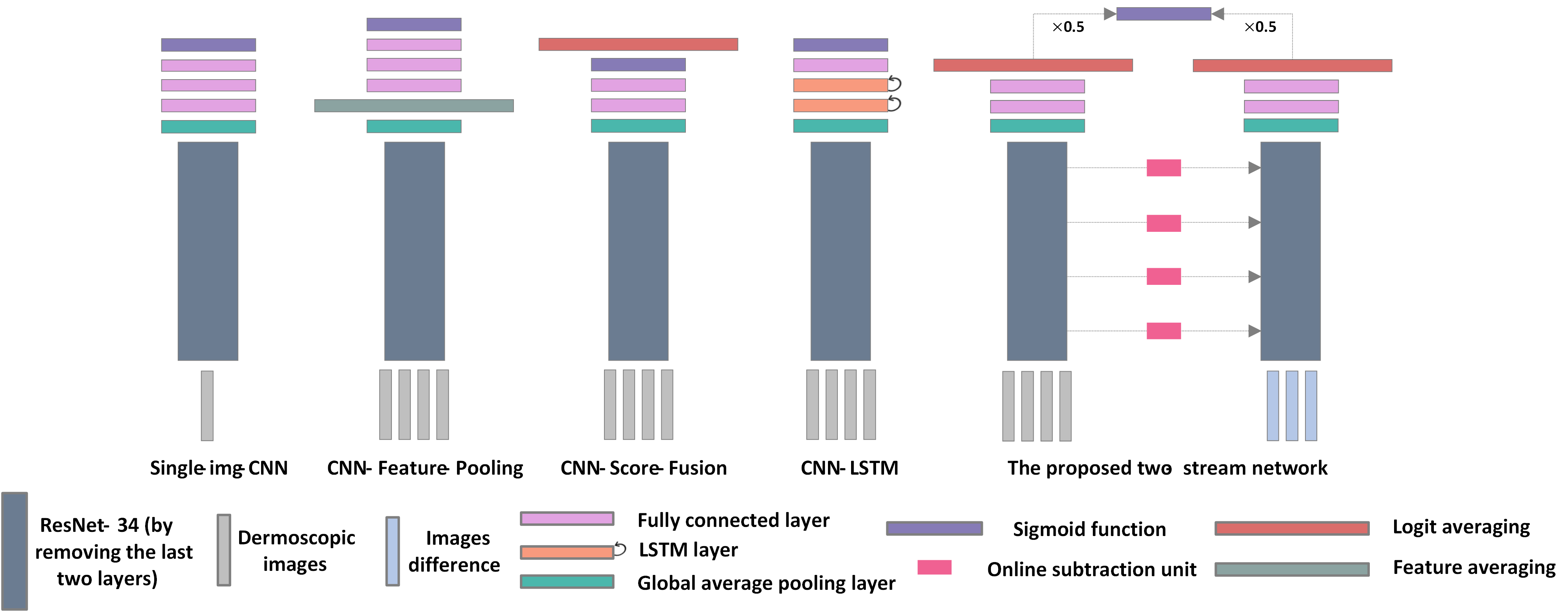}
\caption{The architectures of the proposed model and other comparative models. We omit the dropout layer, batch normalisation layer, or activation layer in the figure for simplicity.} 
\label{fig: cmp_models}
\end{figure}

We provide detailed architecture configurations of the proposed spatio-temporal network and the other comparative models in Table \ref{tab: config_models} and Fig. \ref{fig: cmp_models}. As mentioned above, the spatial stream network and the temporal network of the proposed method share a similar network architecture, and the Single-img-CNN, CNN-Score-Fusion and CNN-Feature-Pooling have the same network configuration. \YZ{All models used the same ImageNet pre-trained ResNet34 as their backbone}. The first three models share a similar network architecture, and the last fully connected (FC) layer is replaced with two new FC layers with a 32 channels and a classification layer. The CNN-LSTM model was built with two LSTM layers with a hidden size of 32 and a classification layer. \YZ{Additionally, we provide the total parameters (para) of each model in Table \ref{tab: config_models}. Because the two sub-networks of ISTN share the same backbone, the total number of parameters is very similar to that of the other two comparative models.}

\begin{table}[ht]
\centering
\caption{Configuration of the proposed model and the other comparative models. Conv, BN, and ReLU denote the convolutional layer, batch normalisation layer, and activation function of the rectified linear unit, respectively.}
\tiny
\begin{tabular}{c|c|c}
\hline
\begin{tabular}[c]{@{}c@{}}\textbf{Single-img-CNN}\\ \textbf{CNN-Score-Fusion}\\ \textbf{CNN-Feature-Pooling}\end{tabular} & \textbf{CNN-LSTM} & \begin{tabular}[c]{@{}c@{}}\textbf{Spatial network} \textbf{\&}\\ \textbf{Temporal network}\\ \textbf{of the ISTN}\end{tabular} \\ \hline \hline
\multicolumn{3}{c}{\begin{tabular}[c]{@{}c@{}}ResNet 34 by removing the last global averaging layer and \\ the fully connected layer\end{tabular}} \\ \hline 
\multicolumn{2}{c|}{Global average pooling} & Dropout, p=0.5 \\ \hline
\multicolumn{2}{c|}{Dropout, p=0.5} & Global average pooling \\ \hline
FC, 512$\times$32 & \begin{tabular}[c]{@{}c@{}}LSTM layer, hidden size=32\\ Dropout, p=0.5\end{tabular} & Dropout, p=0.5 \\ \hline
\begin{tabular}[c]{@{}c@{}}Dropout, p=0.5\\  BN \& ReLU\end{tabular} & \begin{tabular}[c]{@{}c@{}}LSTM layer, hidden size=32\\ Dropout, p=0.5\end{tabular} & \begin{tabular}[c]{@{}c@{}}FC, 512$\times$16\\ BN \& ReLu\end{tabular} \\ \hline
FC, 32$\times$32 & \multirow{4}{*}{FC, 32$\times$1} &  \multirow{4}{*}{FC, 16$\times$1} \\ 
\cline{1-1} \begin{tabular}[c]{@{}c@{}}Dropout, p=0.5\\ BN \& ReLU\end{tabular} &  & \\
\cline{1-1} FC, 32$\times$1 &  &  \\ \hline
\multicolumn{3}{c}{Sigmoid function} \\ \hline
Total para: 21302439 & Total para: 21304289 & Total para: 21317570 \\ \hline
\end{tabular}\label{tab: config_models}
\end{table}

\subsubsection{Training Details}
All experiments were conducted using the Pytorch library. We adopted Adam to optimise the models with a batch size of 32 and an initial learning rate of 0.001. During training, we reduced the learning rate by a factor of five once the validation loss did not decrease within ten epochs. For models initialised with pre-trained network parameters, we froze the mean and variance of all batch normalisation layers to reduce overfitting. Standard data augmentation techniques, such as random resized cropping, colour transformation, and flipping, were used in all experiments. Each dermoscopic image was resized to a fixed size of 320$\times$320 before being input into the models. During the test phase, we utilized ten crop augmentations and then averaged the final predictions.

\subsection{Effect of incorporating temporal information}
\begin{figure}
\centering
\includegraphics[width=0.5\textwidth]{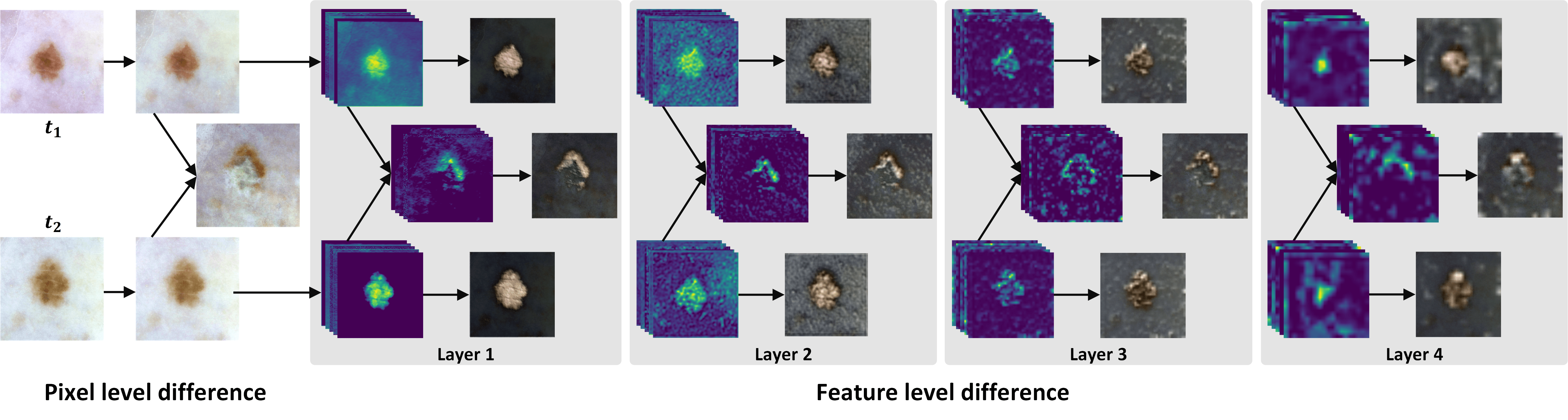}
\caption{Visualisation of the temporal learning process from an example case. To intuitively present the visualisation, we first aggregate the feature difference map to RGB space and then further overlay it to the input image.} \label{fig: diff feat map}
\end{figure}

To better illustrate the temporal difference learning process of the proposed model, we present the pixel-level differences and feature-level differences across various layers on two consecutive images in Fig.~\ref{fig: diff feat map}. We observed that our model successfully captured the new growth region of the lesion. Moreover, the intermediate convolutional activation maps demonstrate that the final prediction is made in the foreground lesion area. 

\end{document}